\documentclass[aps,preprint,onecolumn,secnumarabic,nobalancelastpage,amsmath,amssymb,
nofootinbib]{revtex4}

\RequirePackage{fix-cm}
\usepackage{enumerate}

\usepackage{pst-plot}
\usepackage{pstricks,pstricks-add}
\usepackage[hang,nooneline,scriptsize]{subfigure}

\usepackage{graphicx}      
\usepackage{url}           
\usepackage{dcolumn}
\usepackage{bm}            
\usepackage{mathrsfs}
\usepackage{epstopdf}
\usepackage{color}

\begin{document}

\preprint{APS/123-QED}

\title{Analytical solutions of the geodesic equation in the (rotating) black string-(anti-) de sitter spacetime}

\author{Sobhan Kazempour}
\author{Reza Saffari}%
\email{rsk@guilan.ac.ir}
\author{Saheb Soroushfar}
 \affiliation{Department of Physics, University of Guilan, 41335-1914, Rasht, Iran.
}%

\date{\today}

\begin{abstract}
In this paper we add a compact dimension to Schwarzschild-(anti-) de sitter and Kerr-(anti-) de sitter spacetimes, which describes (rotating) black string-(anti-) de sitter spacetime. We study the geodesic motion of test particles and light rays in this spacetime. We present the analytical solutions of the geodesic equations in terms of Weierstrass elliptic and Kleinian sigma hyperelliptical functions. We also discuss the possible orbits and classify them according to particle's energy and angular momentum. Moreover, the obtained results, are compared to Schwarzschild-(anti-) de sitter and Kerr-(anti-) de sitter spacetimes.
\end{abstract}

\maketitle

\section{INTRODUCTION}

To explain events in cosmic scale, the introduction of a cosmological term  into the Einstein field equation, can be advantageous \cite{Hackmann:2010zz}.
Another discussions on cosmological constant are, its influence on the creation of gravitational waves and the physics of binary systems, which play a significant role in testing General Relativity \cite{Hackmann:2010zz,Naf:2008sf,Barrabes:2007ff}. 
Gravitational waves of massive cosmic events such as merging black holes or neutron star collisions can be generated. Recently lisa pathfinder with Vega rocket from Kuru space-port in French Guiana was launched to detect gravitational waves \cite{Lisa}.
One of the first efforts to unify the electromagnetic and gravity forces, is adding a compact dimension to general relativity \cite{T.K,Klein:1926tv}. Thus, adding a compact dimension to the Schwarzschild--(anti-) de sitter and Kerr--(anti-) de sitter metric, it describes a five-dimensional (rotating) black string-(anti-) de sitter spacetime.
Five-dimensional black holes have special importance. Recently some groups have simulated a Five-dimensional black holes such that they consider the universe as a Five-dimensional black hole and then they study the laws of general relativity on it \cite{Seahra:2003bu,Figueras:2014dta,Figueras:2015hkb}.
The calculation of gravitational waves and also, a systematic study on the last stable spherical and circular orbits, profits from analytical solutions of geodesic equations \cite{Hackmann:2010zz,Dexter:2009fg}.\\
Analytical solution of geodesic equation in Schwarzschild spacetime with use of elliptic functions was demonstrated by Hagihara \cite{Y. Hagihara}.
Moreover, the analytical solution of the equations of motion in the four dimensional Schwarzschild--de Sitter \cite{Hackmann:2008zz}, Kerr--de Sitter \cite{Hackmann:2010zz}, in higher dimensional Schwarzschild, Schwarzschild-(anti-)de Sitter, Reissner-Nordstrom, Reissner Nordstrom--(anti-) de Sitter \cite{Hackmann:2008tu}, and in higher dimensional Myers-Perry spacetimes \cite{Enolski:2010if}, were presented. \\
Also, the geodesics equations were solved analytically, in the singly spinning \cite{Grunau:2012ai}, (charged) doubly spinning black ring \cite{Grunau:2012ri}, and (rotating) black string \cite{Grunau:2013oca}, Schwarzschild and Kerr pierced by black string spacetimes \cite{Hackmann:2009rp,Hackmann:2010ir}. Moreover, geodesic motion in the spacetime of BTZ and GMGHS black holes and also black holes in conformal gravity and $f(R)$ gravity, were investigated analyticlly in Refs.\cite{Soroushfar:2015dfz,Soroushfar:2016yea,Hoseini:2016nzw,Soroushfar:2015wqa,Soroushfar:2016azn,Soroushfar:2016esy}.\\
In this paper we add a compact dimension to Schwarzschild-(anti-) de sitter and Kerr-(anti-) de sitter spacetimes, which describes (rotating) black string-(anti-) de sitter spacetime. We observe new behaviour with adding a compact dimension and analysed it in detail. We present the analytical solutions of the geodesic equations in terms of Weierstrass elliptic and Kleinian sigma hyperelliptical functions and discuss about their orbits. First we study the static black string-(anti-) de sitter and we compare it to the Schwarzschild-(anti-) de sitter black hole,and then in the second part, we analyse the rotating black string-(anti-) de sitter and also we compare it to the Kerr-(anti-) de sitter black hole.

\section{black string-(anti) de sitter spacetime}
In this section, we will study the geodesics in the static black string-(anti-) de sitter spacetime and introduce analytical solutions of the equations of motion and orbit types. A detailed analysis of the geodesics in the original Schwarzschild-(anti-) de sitter spacetime without a compact dimension can be found in e.g.\cite{Hackmann:2008zz}.

\subsection{The geodesic equations} 
If we add an extra compact spatial dimension $w$ to Schwarzschild-(anti-) de sitter spacetime, it takes this form:
\begin{align}\label{ds1}
ds^{2}=-(1-2\frac{M}{r}-\frac{\Lambda r^{2}}{3})d t^{2}+(\dfrac{1}{1-2\frac{M}{r}-\frac{\Lambda r^{2}}{3}})d r^{2}+r^{2}(d\theta^{2}+\sin^2\theta d\varphi^{2})+dw^{2}.
\end{align}
This black string-(anti) de Sitter spacetime is characterized
by the Schwarzschild-radius $r_{S}=2M$ related
to the mass $M$ of the gravitating body, and the cosmological
constant $\Lambda$. The geodesic equation has to be completed
by the normalization condition $g_{\mu\nu}\frac{dx^{\mu}}{ds}\frac{dx^{\nu}}{ds}=\varepsilon$, where for massive particles $\varepsilon=1$ and for light $\varepsilon=0$.\\
The Hamilton-Jacobi equation 
\begin{equation}\label{Hamilton0}
\dfrac{\partial S}{\partial\tau}+\frac{1}{2}\ g^{\mu\nu}\dfrac{\partial S}{\partial X^{\mu}}\dfrac{\partial S}{\partial X^{\nu}}=0.
\end{equation}
can be solved with an ansatz for the action
\begin{equation}
\label{S0}
S=\frac{1}{2}\varepsilon \tau - Et+L\varphi +Jw+ S_{r} (r) ,
\end{equation}
Where, $E$, is the energy, $L$, denotes the angular momentum and $J$, is a new constant of motion according to the compact dimension $w$, and $\tau$, is an affine parameter along the geodesic. We set $\theta=\frac{\pi}{2}$, since the orbits lie in a plane due to the spherical symmetry of the original Schwarzschild-(anti) de Sitter metric.\\
Using Eqs.~(\ref{ds1})--(\ref{S0}) and with the help of the mino time $\lambda$ as $r^{2}d\lambda=d\tau$ \cite{Mino:2003yg} we get
\begin{align}\label{drdla}
(\frac{dr}{d\lambda})^{2}=-r^{4}(1-2\frac{M}{r}-\frac{\Lambda r^{2}}{3})(\varepsilon +J^2)+r^{4}E^{2}-L^{2}r^{2}(1-2\frac{M}{r}-\frac{\Lambda r^{2}}{3})=R(r),
\end{align}
\begin{align}\label{J0}
(\frac{dw}{d\lambda})=r^{2}J,
\end{align}
\begin{align}\label{dphi0}
(\frac{d\varphi}{d\lambda})=L,
\end{align}
\begin{align}\label{dt0}
(\frac{dt}{d\lambda})=\dfrac{r^{2}E}{(1-\frac{2M}{r}-\frac{\Lambda r^{2}}{3})}.
\end{align}
For $J=0$, these equations, are the same as in the original Schwarzschild-(anti-) de Sitter spacetime without the compact dimension \cite{Hackmann:2008zz}.
\\
For simplicity, we rescale the parameters appearing in eqs.~(\ref{drdla})--(\ref{dt0}), with following dimensionless parameters
\begin{align}
\tilde{r}=\dfrac{r}{M} , \qquad \tilde{t}=\dfrac{t}{M} , \qquad \tilde{L}=\dfrac{L}{M},\qquad \tilde{\Lambda}=\frac{1}{3}\Lambda M^{2}, \qquad \tilde{w}=\dfrac{w}{M}, \qquad \gamma =M\lambda .
\end{align}
Then, the equations~(\ref{drdla})--(\ref{dt0}) can be rewritten as
\begin{align}\label{drdla'}
(\frac{d\tilde{r}}{d\gamma})^{2}=-\tilde{r}^{4}(1-\frac{2}{\tilde{r}}-\tilde{\Lambda} \tilde{r}^{2})(\varepsilon +J^2)+\tilde{r}^{4}E^{2}-\tilde{L}^{2}\tilde{r}^{2}(1-\frac{2}{\tilde{r}}-\tilde{\Lambda} \tilde{r}^{2})=\tilde{R}(\tilde{r}),
\end{align}
\begin{align}\label{J0'}
(\frac{d\tilde{w}}{d\gamma})=\tilde{r}^{2}J,
\end{align}
\begin{align}\label{dphi0'}
(\frac{d\varphi}{d\gamma})={\tilde{L}},
\end{align}
\begin{align}\label{dt0'}
(\frac{d\tilde{t}}{d\gamma})=\dfrac{\tilde{r}^{2}E}{(1-\frac{2}{\tilde{r}}-\tilde{\Lambda} \tilde{r}^{2})}.
\end{align}
Equation~(\ref{drdla'}) proposes the introduction of an effective
potential
\begin{align}
V_{eff}=-\dfrac{(\tilde{\Lambda} \tilde{r}^{3}-\tilde{r}+2)(J^{2}\tilde{r}^{2}+\varepsilon \tilde{r}^{2}+\tilde{L}^{2})}{\tilde{r}^{3}}.
\end{align}
The plots of this effective potential are shown in Figs.\ref{11} - \ref{22}

\subsection{Types of radial motion}
\begin{figure}[!ht]
\centering
\subfigure[ ]{
\includegraphics[width=6cm]{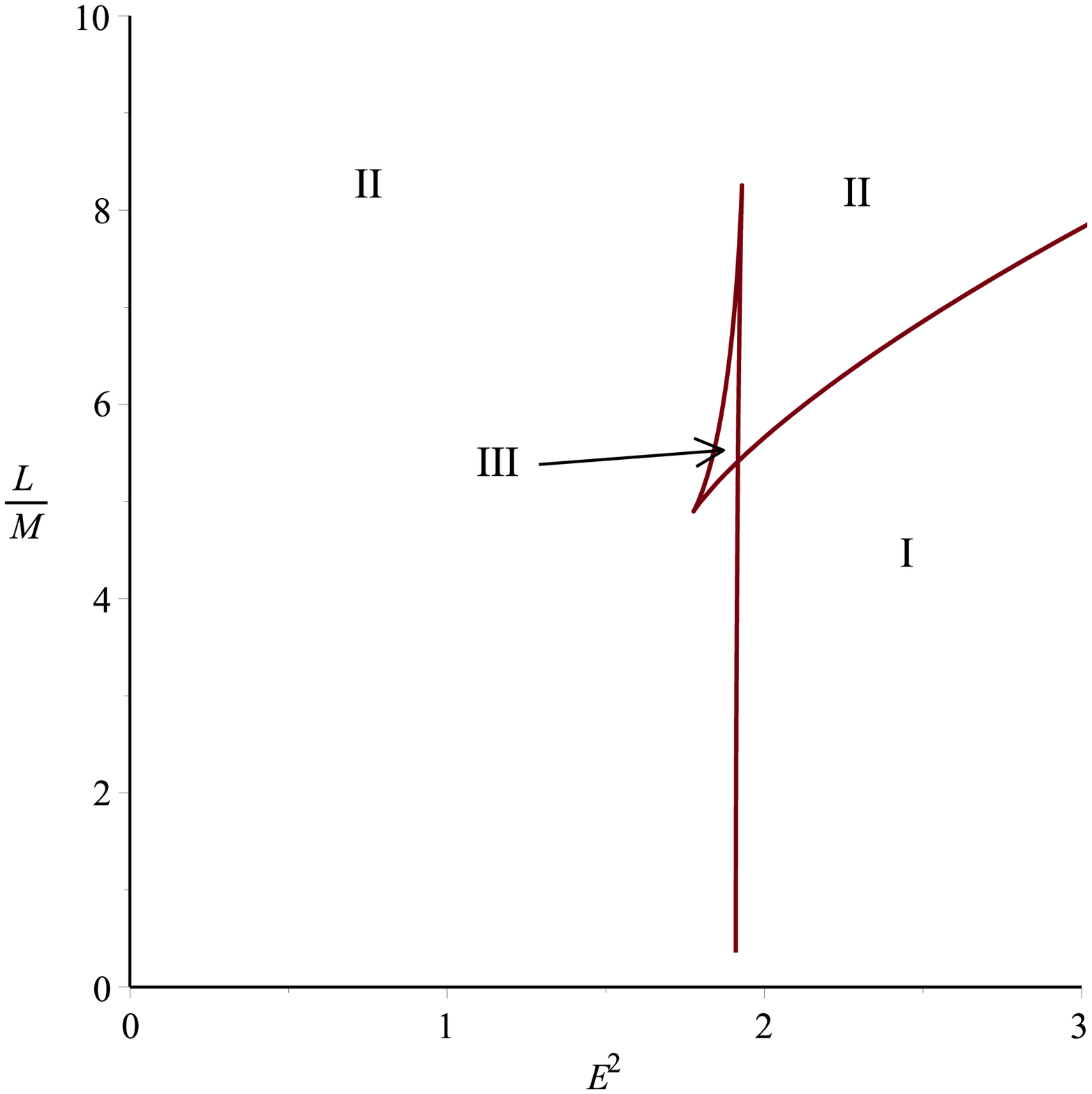}
	\label{1}}
\subfigure[  ]{
\includegraphics[width=6cm]{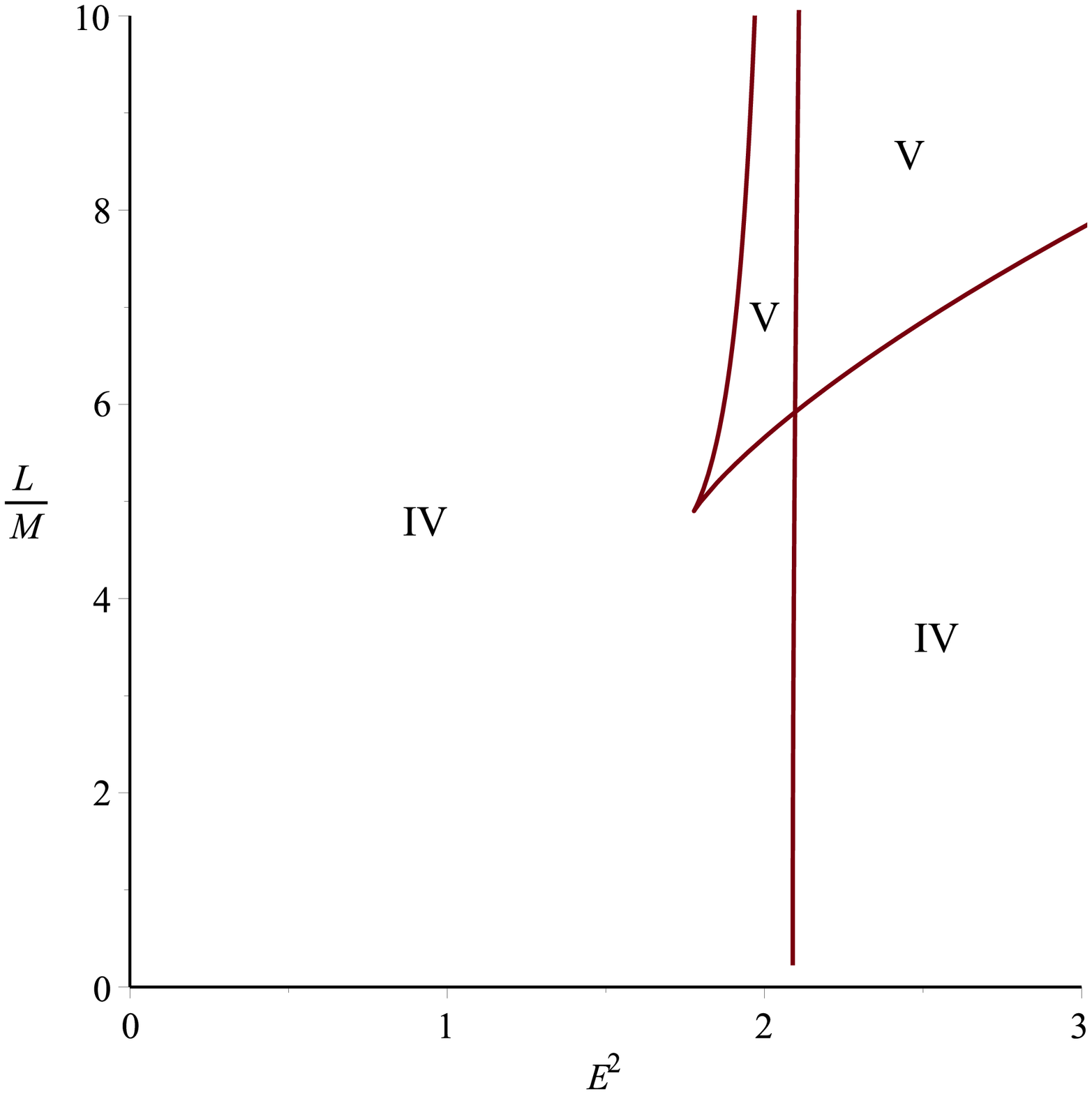}
	\label{2}}
\caption[figs]
{ Regions of different types of geodesic motion in black string-(anti) de sitter for $\varepsilon=1$, $J=1$ \subref{1} $\Lambda=\frac{1}{3}\times 10^{-5}$, 
 \subref{2} $\Lambda=-\frac{1}{3}\times10^{-5}$.}
\label{3}
\end{figure}

\begin{figure}[!ht]
\centering
\subfigure[ ]{
\includegraphics[width=6cm]{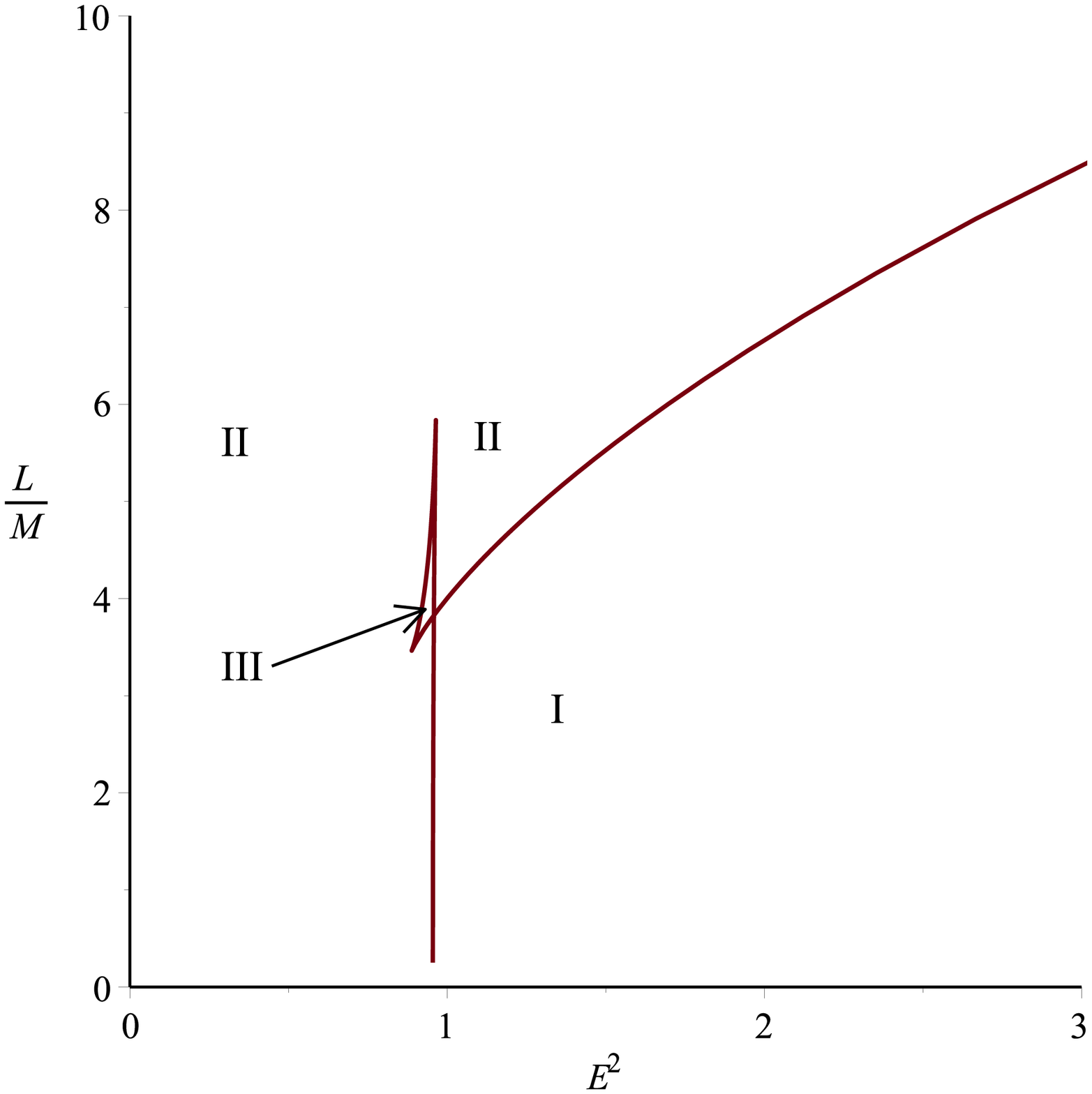}
	\label{4}}
\subfigure[  ]{
\includegraphics[width=6cm]{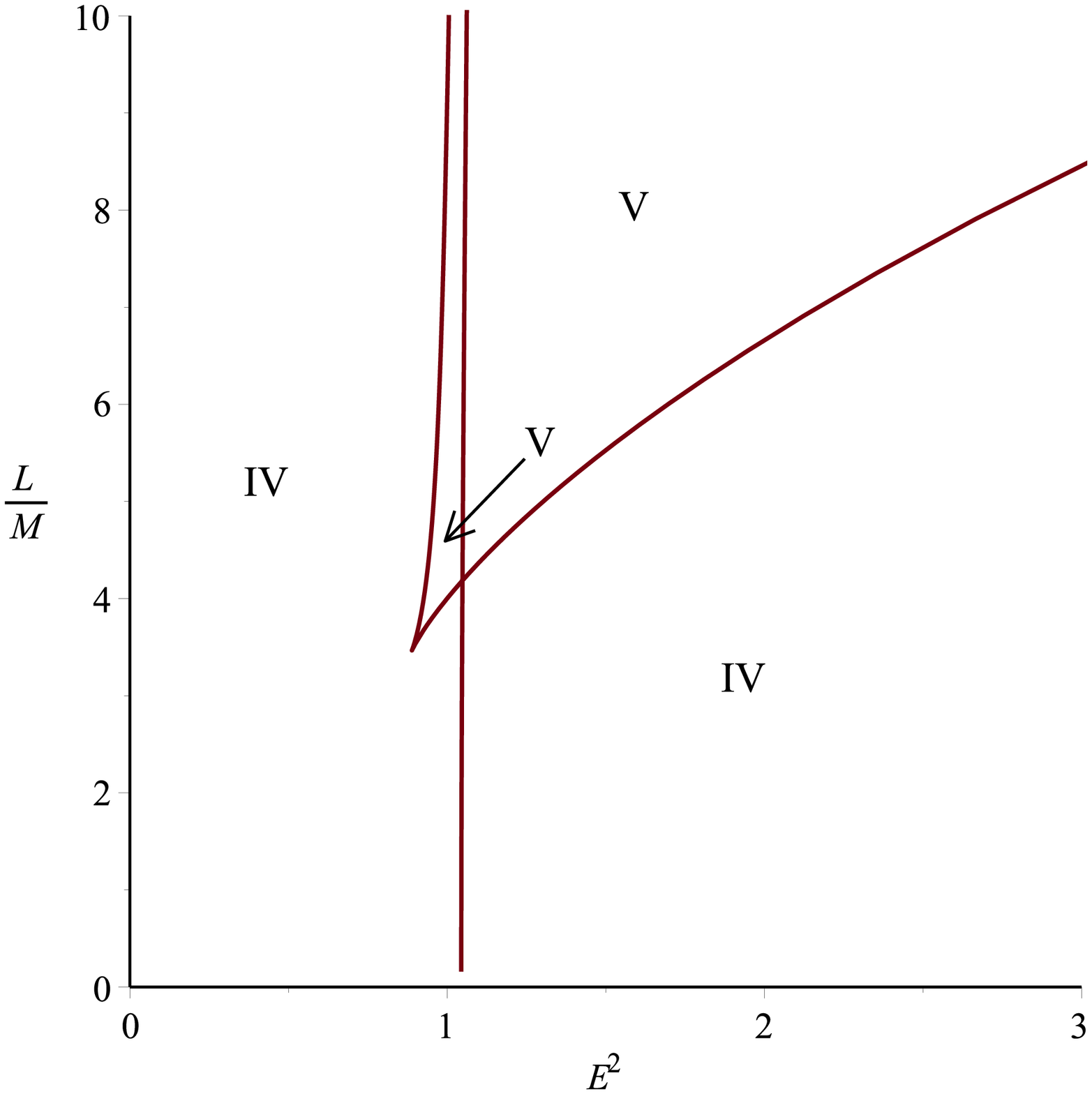}
	\label{5}}
\caption[figs]
{ Regions of different types of geodesic motion in black string-(anti) de sitter for $\varepsilon=0$, $J=1$ \subref{4} $\Lambda=\frac{1}{3}\times 10^{-5}$, 
 \subref{5} $\Lambda=-\frac{1}{3}\times10^{-5}$.}
\label{6}
\end{figure}

We rewrite eq.~(\ref{drdla'}) as
\begin{align}
\tilde{R}(\tilde{r})=(\tilde{\Lambda} J^{2}+\tilde{\Lambda} \varepsilon)\tilde{r}^{6}+(\tilde{L}^{2}\tilde{\Lambda} +E^{2}-J^{2}-\varepsilon)\tilde{r}^{4}+(2 J^{2}+2\varepsilon)\tilde{r}^{3}-\tilde{L}^{2}\tilde{r}^{2}+2\tilde{L}^{2}\tilde{r},
\end{align}
and use this equation to determine the possible types of orbit.
The zeros of the polynomial $\tilde{R}(\tilde{r})$ are turning points of orbits of light and test particles.\\
Using $\tilde{R}(\tilde{r})=0$, $\dfrac{d\tilde{R}(\tilde{r})}{d\tilde{r}}=0$, conditions, we plot $\tilde{L}-E^{2}$, diagrams for timelike ($\varepsilon =1$), and null geodesics ($\varepsilon =0$),
which are shown in Figs. \ref{3} and \ref{6}. For $J=0$, these figures, are the same as in the Schwarzschild-(anti-) de Sitter spacetime figures \cite{Hackmann:2008zz}.
\\
The list of all possible orbits are demonstrated in the spacetimes described by the metric Eq.~(\ref{ds1}):
\begin{enumerate}
\item Escape orbit ($EO$) with range $\tilde{r}$ $\in$ [$r_{1}$,$\infty$) with $r_{1}>\tilde{r}_{+}$, or with range $\tilde{r}$ $\in$ ($-\infty$ ,$ r_{1}$] with $r_{1}<0$.
\item Bound orbit ($BO$) with range $\tilde{r} \in [r_{1}, r_{2}]$ with $0 < r_{1} < r_{2}$ and. (a) either $r_{1},r_{2}>r_{+}$ or (b) $r_{1},r_{2}<r_{-}.$
\item Terminating orbit ($TO$) with ranges $\tilde{r} \in [0,\infty)$ or $\tilde{r} \in [0, r_{1}]$ with. (a) either $r_{1}\geqslant \tilde{r}_{+}$ or (b) $0<r_{1}<\tilde{r}_{-}$.
\end{enumerate}
\newpage

Five different regions of geodesic motion for $\tilde{\Lambda}=\pm10^{-5}$ and $J=1$, can be identified:
\begin{enumerate}
\item ~Region I: $\tilde{R}(\tilde{r})$ has $0$ positive real zeros and $\tilde{R}(\tilde{r})\geqslant 0$  for $0\leqslant \tilde{r}$. Possible orbit types:~terminating orbits. 
\item ~Region II: $\tilde{R}(\tilde{r})$ has $2$ positive real zeros $r_{1}< r_{2}$ with $\tilde{R}(\tilde{r}) > 0$ for $0\leqslant \tilde{r}\leqslant r_{1}$ and $r_{2}\leqslant\tilde{r}$. Possible orbit types: escape and terminating orbits.
\item ~Region III: $\tilde{R}(\tilde{r})$ has $4$ positive real zeros $r_{i} < r_{i+1}$ with $\tilde{R}(\tilde{r}) \geqslant 0$ for $0 \leqslant \tilde{r}\leqslant r_{1}$, $r_{2} \leqslant \tilde{r}\leqslant r_{3}$, and $r_{4}\leqslant \tilde{r}$. Possible orbit types: escape, bound, and terminating orbits
\item ~Region IV: $\tilde{R}(\tilde{r})$ has $1$ positive real zero $r_{1}$ with $\tilde{R}(\tilde{r})\geqslant  0$ for positive $r$. Possible orbit types: terminating orbits.
\item ~Region V: $\tilde{R}(\tilde{r})$ has $3$ positive real zeros $r_{1} < r_{2} < r_{3}$ with $\tilde{R}(\tilde{r})\geqslant 0$ for $0\leqslant \tilde{r}\leqslant r_{1}$ and
$r_{2}\leqslant \tilde{r}\leqslant r_{3}$. Possible orbit types: bound and terminating orbits.
\end{enumerate}

\begin{figure}[!ht]
\centering
\subfigure[ ]{
\includegraphics[width=7cm]{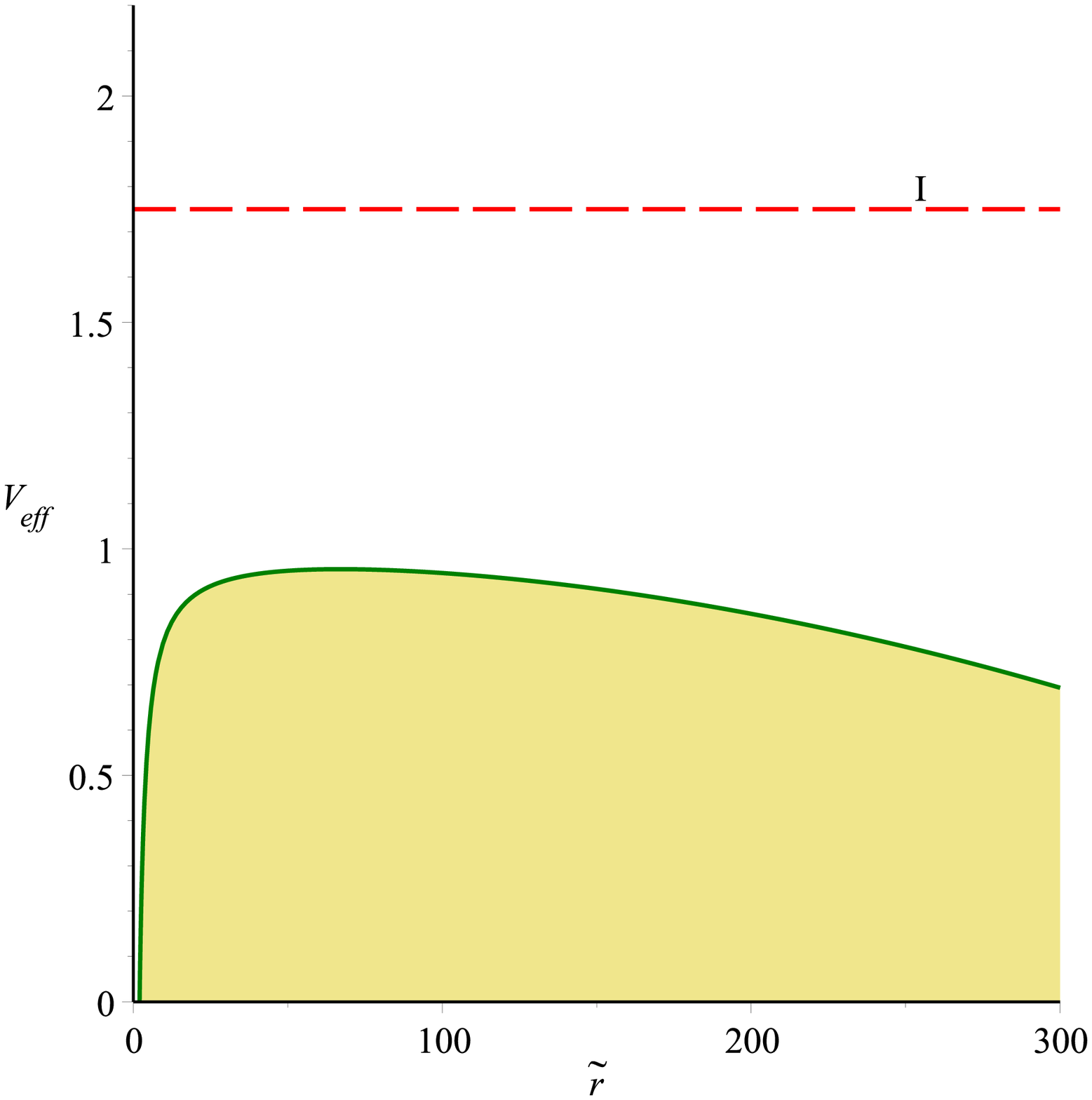}
	\label{7}}
\hspace*{0mm}
\subfigure[  ]{
\includegraphics[width=7cm]{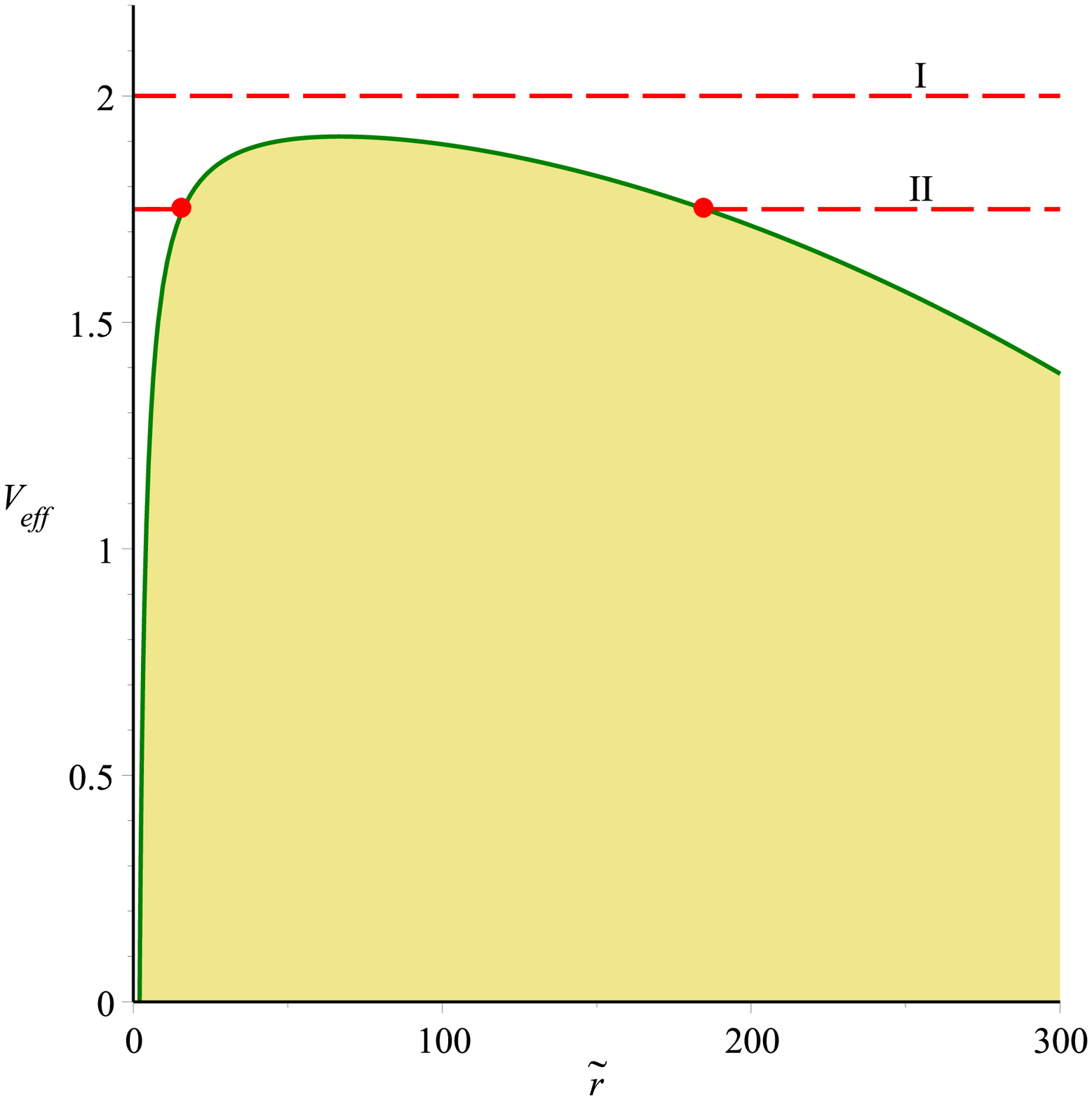}
	\label{8}}
\subfigure[ ]{
\includegraphics[width=7cm]{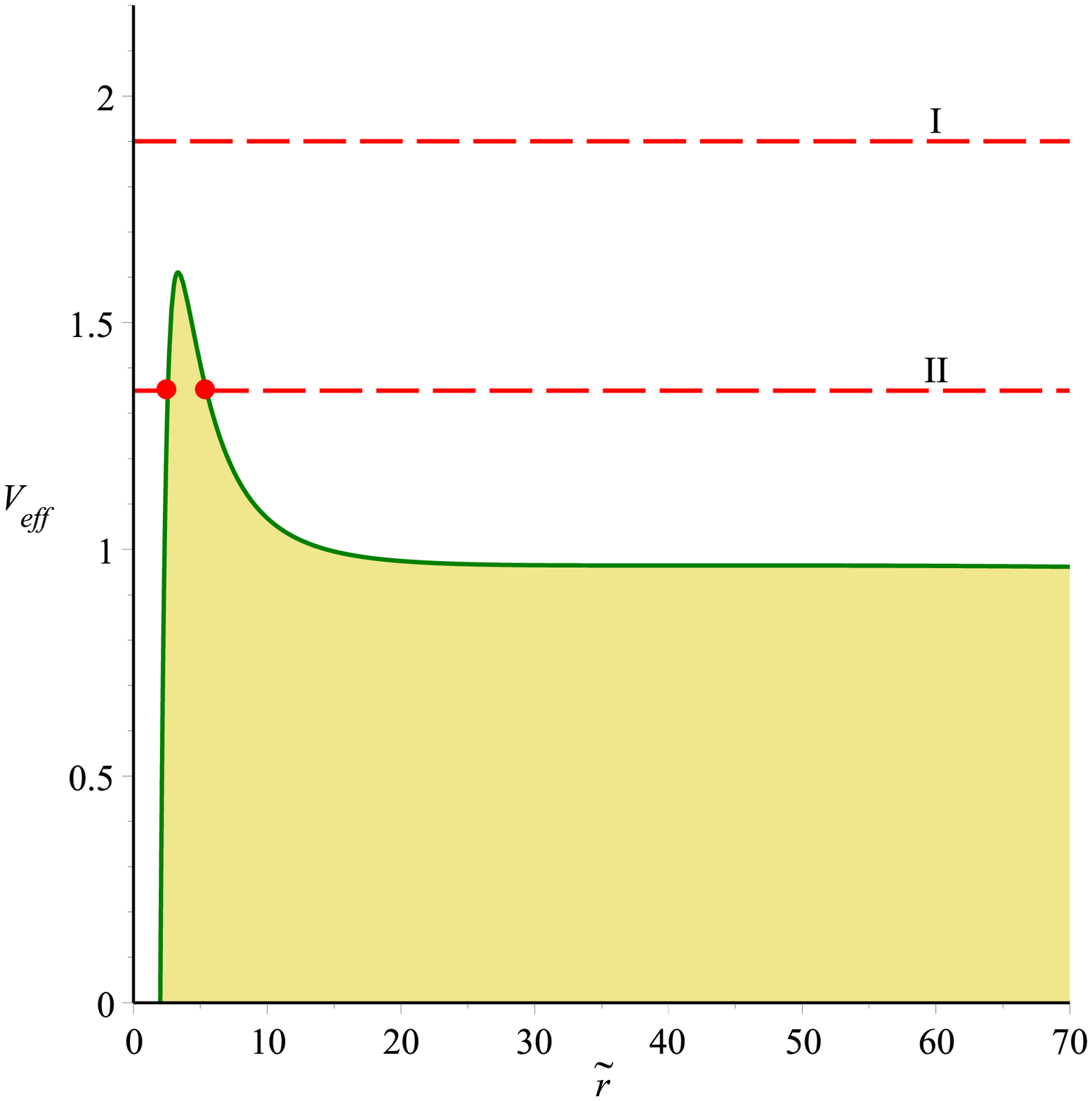}
	\label{9}}
\hspace*{0mm}
\subfigure[  ]{
\includegraphics[width=7cm]{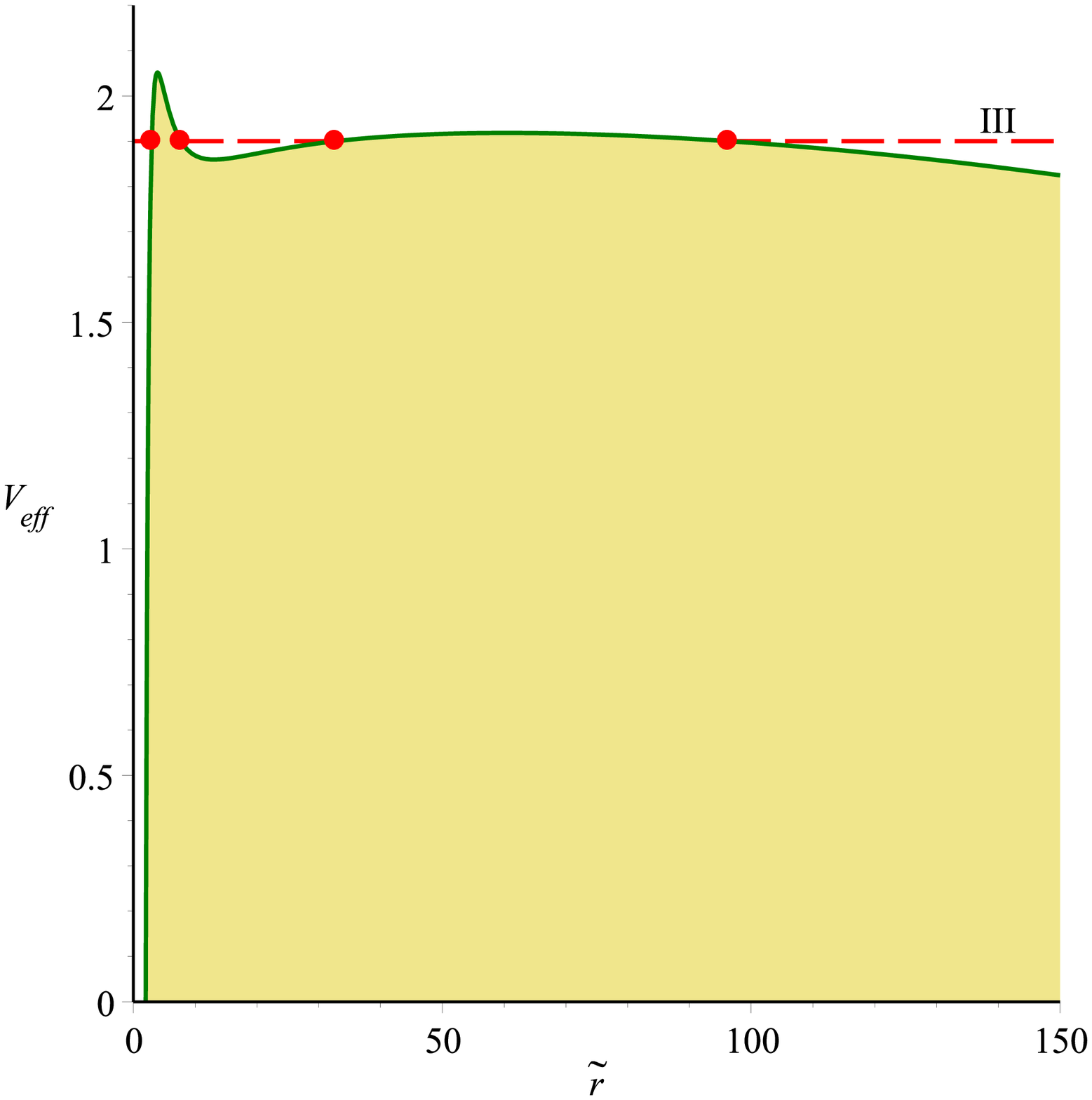}
	\label{10}}

\caption[figs]
{ Effective potentials for test particles ($\varepsilon =1$). The green curves represent to the effective potential. 
The red dashed lines denote the energy. The red dots mark the zeros of the polynomial $R$, which are the turning points of the orbits.
In the khaki area no motion is possible since $\tilde{R}<0$.
Schwarzschild-(anti) de sitter space-time ($J=0$), with $\tilde{L}=\frac{1}{9}$, $\tilde{L}=5.8$, $\tilde{\Lambda}=\frac{1}{3}\times 10^{-5}$,  for \subref{7}, \subref{9}. 
black string-(anti) de sitter space-time ($J=1$), with $\tilde{L}=\frac{1}{9}$, $\tilde{L}=5.8$, $\tilde{\Lambda}=\frac{1}{3}\times 10^{-5}$,  for \subref{8}, \subref{10}.}
\label{11}
\end{figure}

\begin{figure}[!ht]
\centering
\subfigure[ ]{
\includegraphics[width=7cm]{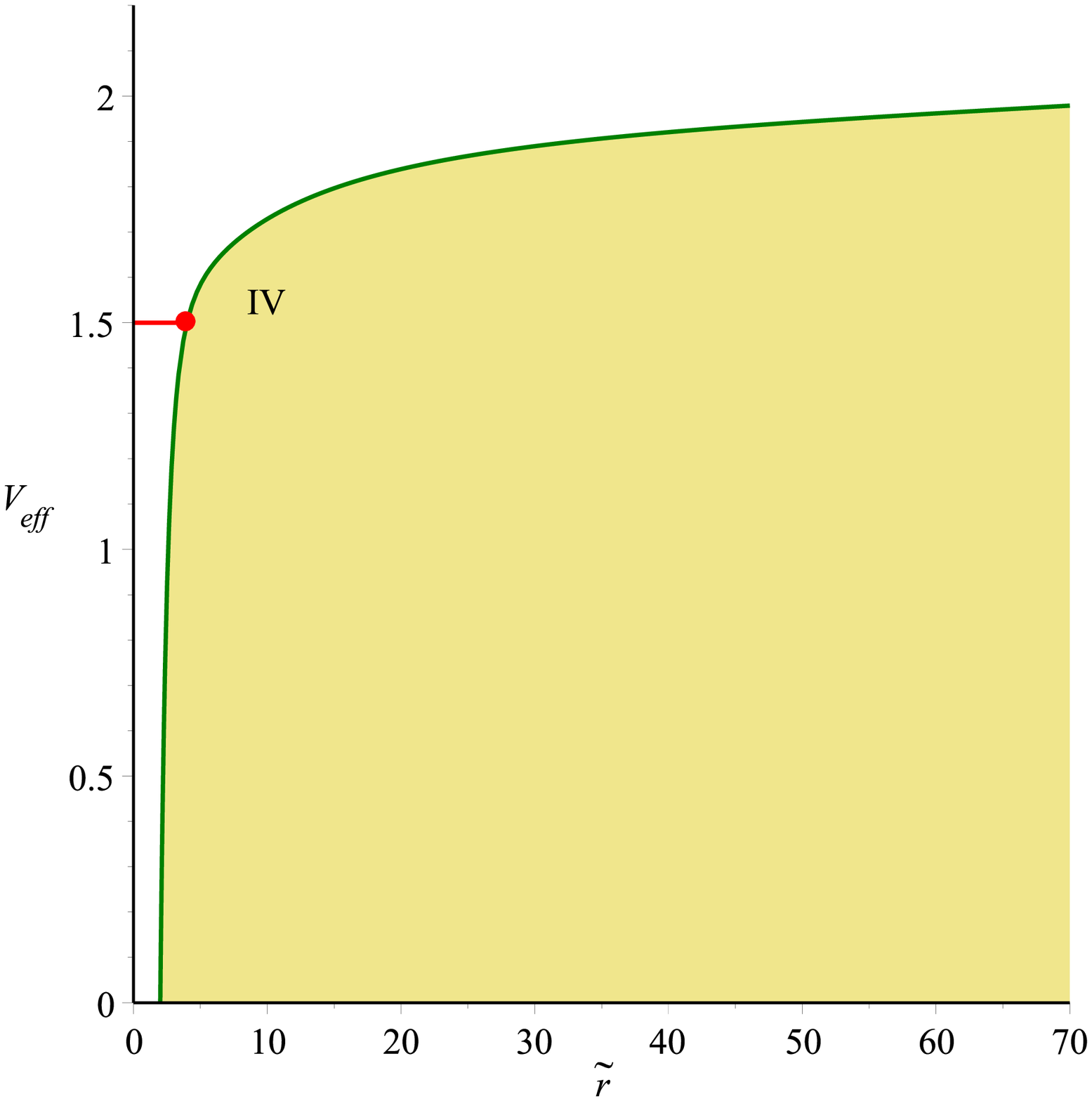}
	\label{12}}
\hspace*{0mm}
\subfigure[  ]{
\includegraphics[width=7cm]{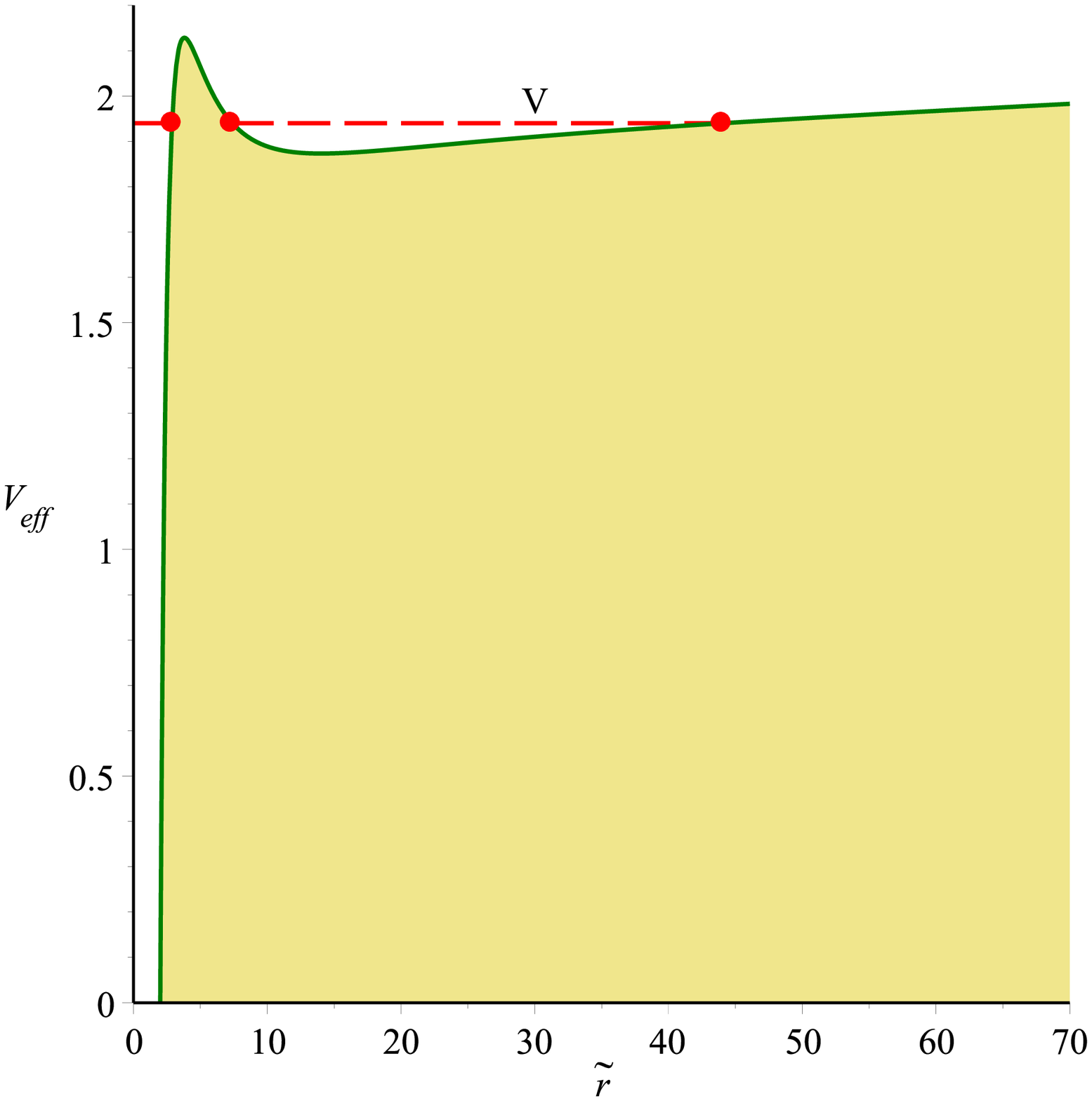}
	\label{13}}
\caption[figs]
{ Effective potentials for test particles ($\varepsilon =1$). The green curves represent to the effective potential. 
The red dashed lines denote the energy. The red dots mark the zeros of the polynomial $R$, which are the turning points of the orbits.
In the khaki area no motion is possible since $\tilde{R}<0$. \subref{12} with $\tilde{L}=4$, $\tilde{\Lambda}=-\frac{1}{3}\times 10^{-5}$, $J=1$, \subref{13} with $\tilde{L}=6$, $\tilde{\Lambda}=-\frac{1}{3}\times 10^{-5}$ ,$J=1$.}
\label{14}
\end{figure}

\begin{figure}[!ht]
\centering
\subfigure[ ]{
\includegraphics[width=7cm]{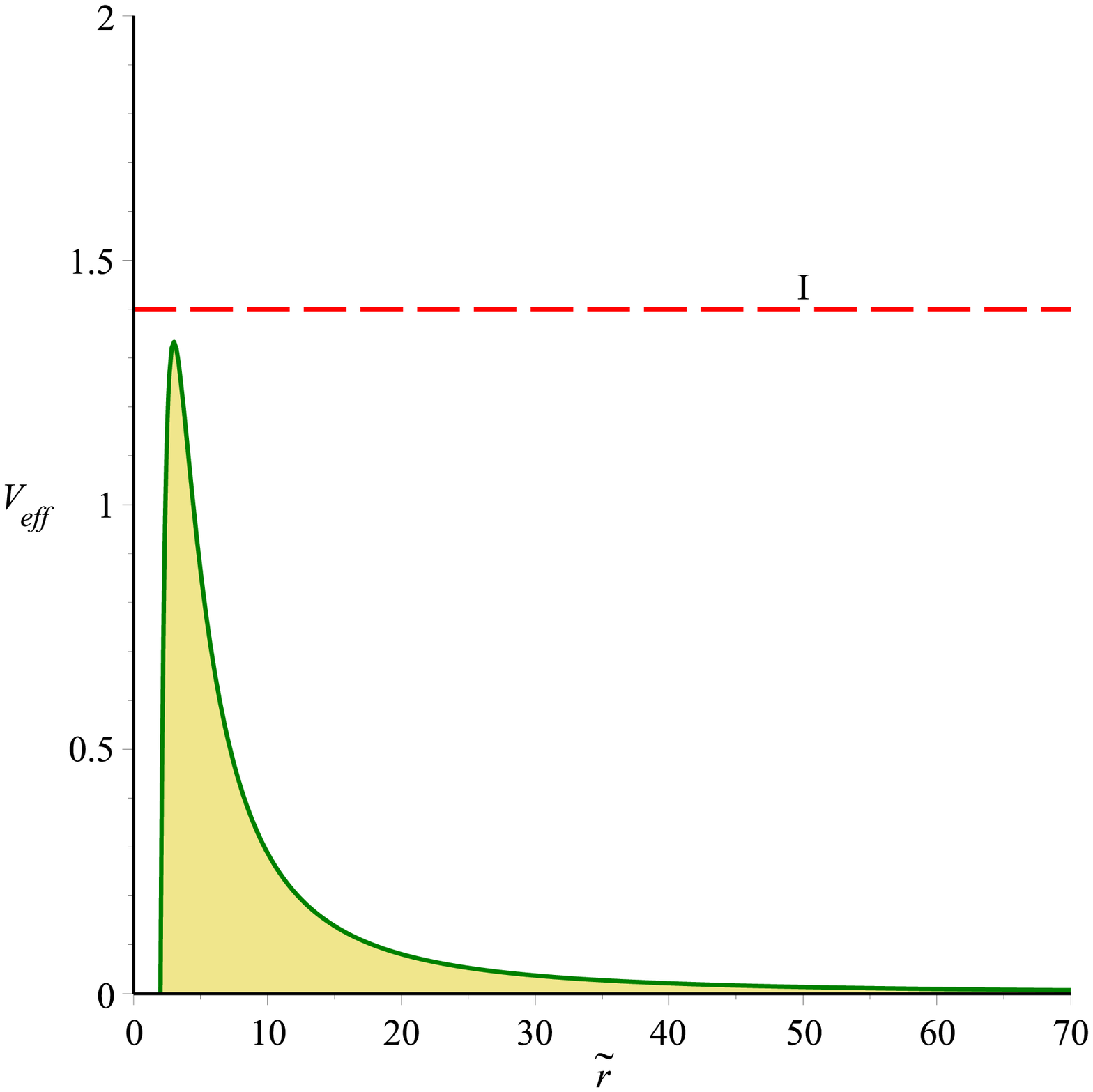}
	\label{151}
}
\hspace*{0mm}
\subfigure[  ]{
\includegraphics[width=7cm]{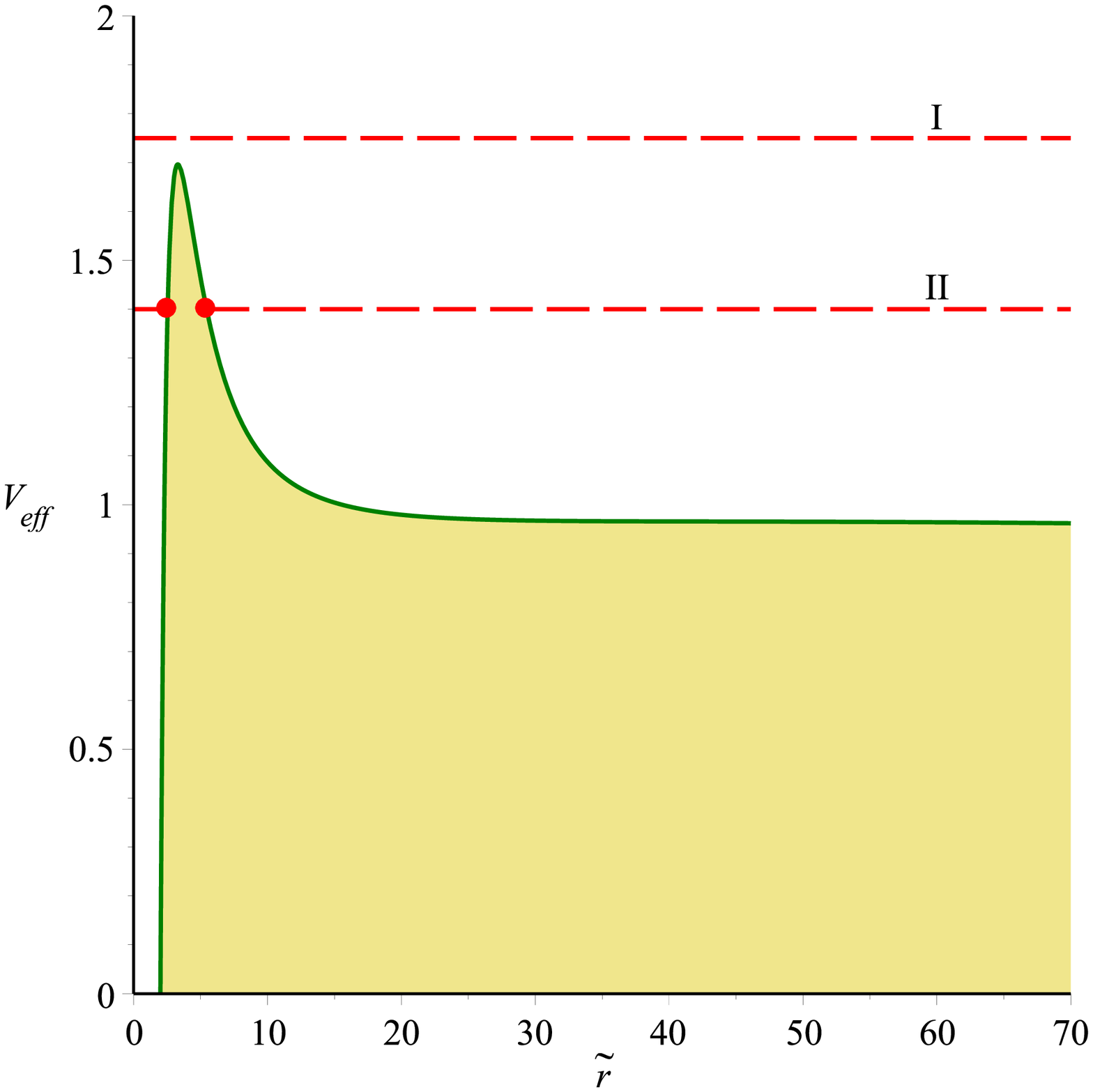}
	\label{161}
}
\subfigure[ ]{
\includegraphics[width=7cm]{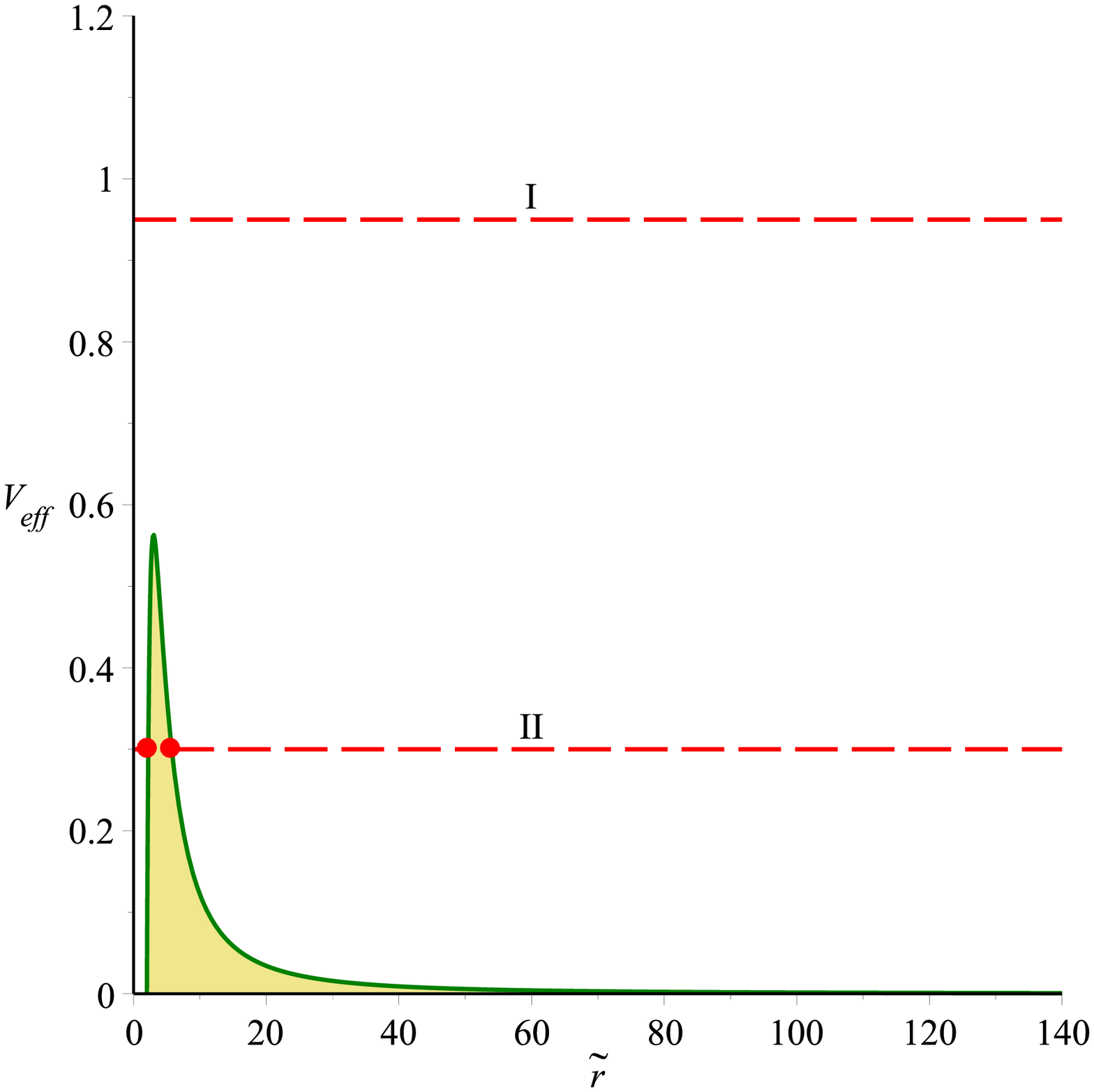}
	\label{171}
}
\hspace*{0mm}
\subfigure[  ]{
\includegraphics[width=7cm]{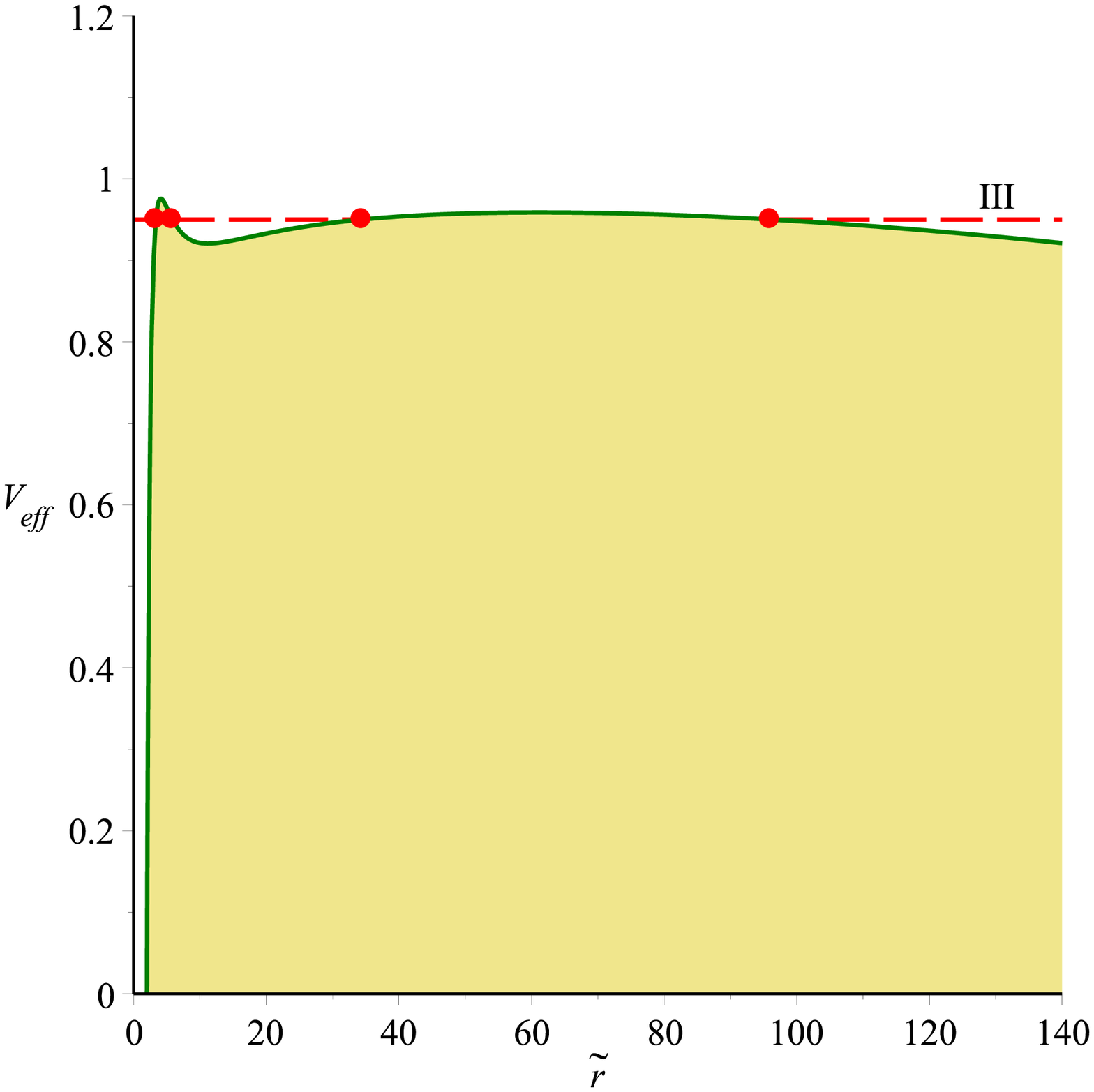}
	\label{18}
}
\caption[figs]
{ Effective potentials for light rays ($\varepsilon =0$). The green curves represent to the effective potential. 
The red dashed lines denote the energy. The red dots mark the zeros of the polynomial $R$, which are the turning points of the orbits.
In the khaki area no motion is possible since $\tilde{R}<0$. 
\subref{151}, \subref{171}, with $\tilde{L}=6$, $\tilde{L}=3.9$, $\tilde{\Lambda}=\frac{1}{3}\times 10^{-5}$, denote  Schwarzschild-(anti) de sitter space-time ($J=0$).
\subref{161}, \subref{18}, with $\tilde{L}=6$, $\tilde{L}=3.9$, $\tilde{\Lambda}=\frac{1}{3}\times 10^{-5}$, denote black string-(anti) de sitter space-time ($J=1$).}
\label{19}
\end{figure}

\begin{figure}[!ht]
\centering
\subfigure[ ]{
\includegraphics[width=7cm]{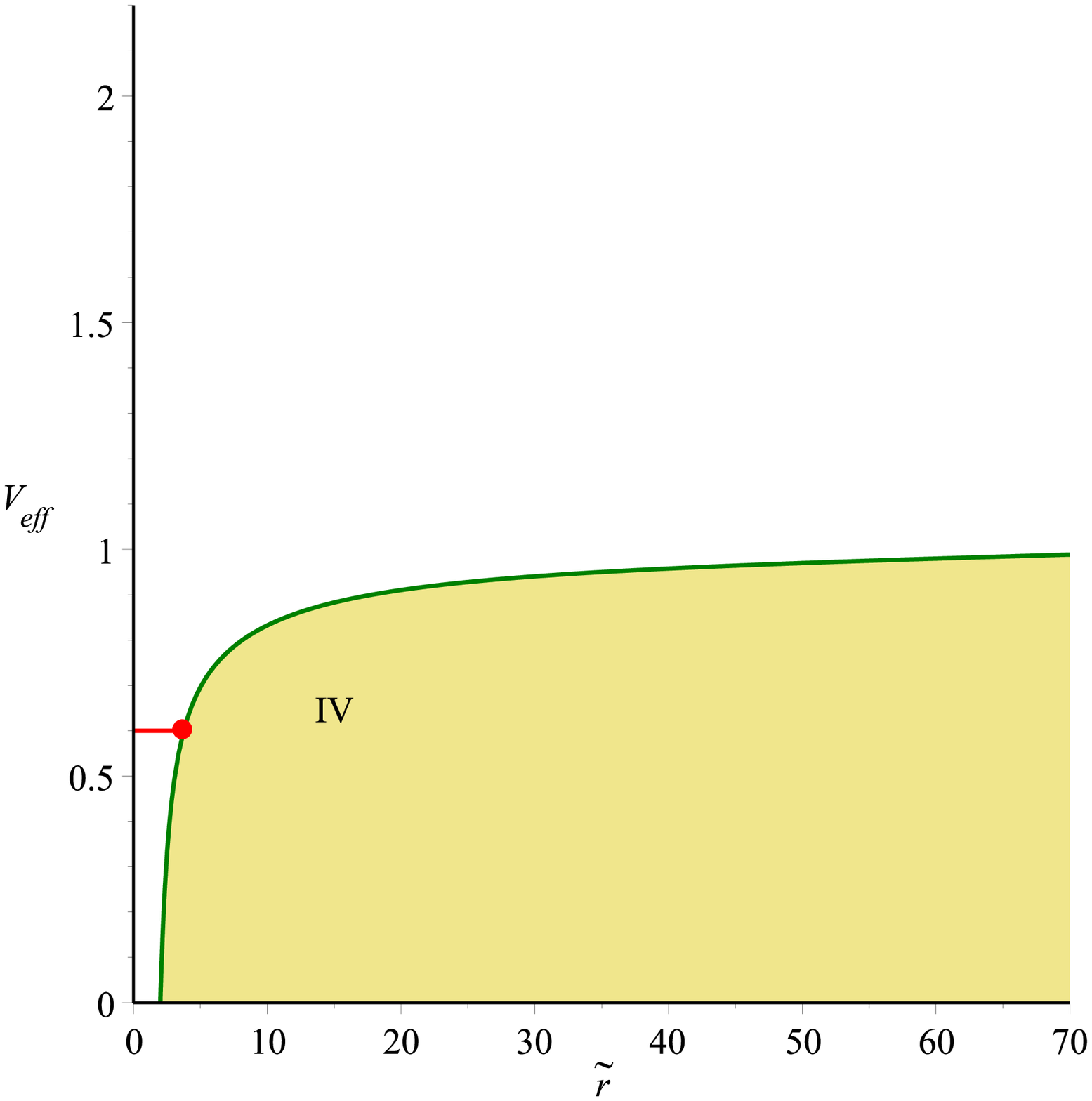}
	\label{20}}
\hspace*{0mm}
\subfigure[  ]{
\includegraphics[width=7cm]{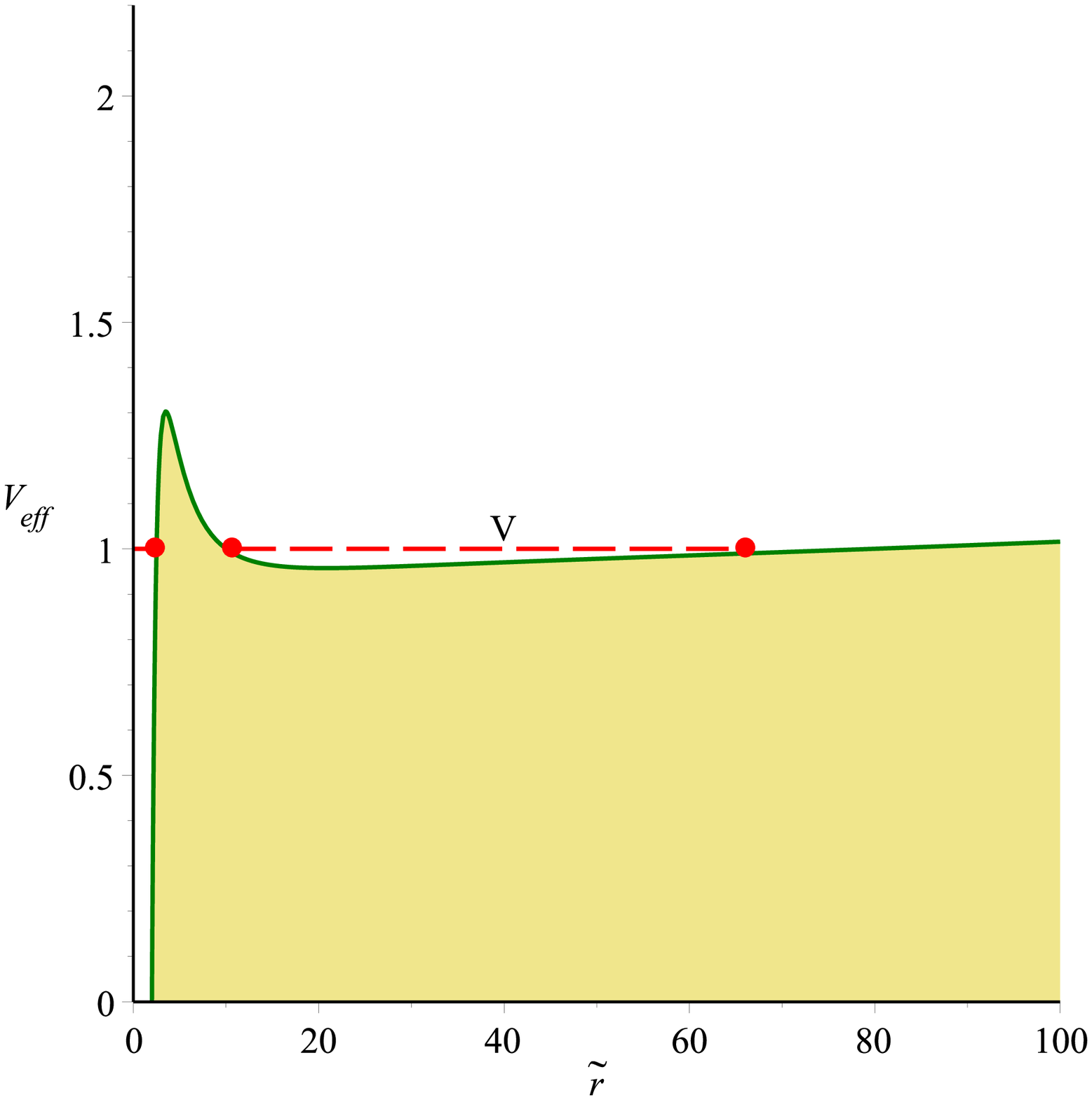}
	\label{21}}
\caption[figs]
{ Effective potentials for test light rays ($\varepsilon =0$). The green curves represent to the effective potential. 
The red dashed lines denote the energy. The red dots mark the zeros of the polynomial $R$, which are the turning points of the orbits.
In the khaki area no motion is possible since $\tilde{R}<0$. \subref{20} with $\tilde{L}=2$, $\tilde{\Lambda}=-\frac{1}{3}\times 10^{-5}$, $J=1$, \subref{21} with $\tilde{L}=5$, $\tilde{\Lambda}=-\frac{1}{3}\times 10^{-5}$ ,$J=1$.}
\label{22}
\end{figure}

For each regions, examples of effective potentials are demonstrated in Figs \ref{11}--~\ref{22}. Also, summary of possible orbit types can be found in Tables~\ref{tab:BSd.orbits} and \ref{tab:BSd.orbits1}.

\begin{table}[!ht]
\begin{center}
\begin{tabular}{|l|l|c|l|}
\hline
Region & Positive zeros & Range of $\tilde{r}$ &  Types of orbits \\
\hline\hline
I & 0 &
$|$$\textbf{--------------------------------}$

& TO
\\  \hline
II & 2 &
$|$$\textbf{---------}$$\bullet$$----$$\bullet$$\textbf{-----------}$ 
  & TO, EO
\\ \hline
III & 4 &
$|$$\textbf{--}$$\bullet$$--$$\bullet$$\textbf{---}$$\bullet$$----$$\bullet$$\textbf{--------}$
  & TO, BO, EO
\\ \hline
\end{tabular}
\caption{Orbit types of a black string-de sitter for 
$J=1$ and $ \tilde{\Lambda}=\frac{1}{3}\times10^{-5}$. The range of the orbits is represented by lines. The
dots show the turning points of the orbits.
 The single vertical line indicates $\tilde{r}=0$.}
\label{tab:BSd.orbits}
\end{center}
\end{table}
\begin{table}[!ht]
\begin{center}
\begin{tabular}{|l|l|c|l|}
\hline
Region & Positive zeros & Range of $\tilde{r}$ &  Types of orbits \\
\hline\hline
IV & 1 &
$|$$\textbf{--}$$\bullet$$------------$

& TO
\\  \hline
V & 3 &
$|$$\textbf{--}$$\bullet$$--$$\bullet$$\textbf{---}$$\bullet$$--------$
  & TO, BO
\\ \hline
\end{tabular}
\caption{Orbit types of a black string-anti de sitter for 
$J=1$ and $ \tilde{\Lambda}=-\frac{1}{3}\times10^{-5}$. The range of the orbits is represented by lines. The
dots show the turning points of the orbits.
 The single vertical line indicates $\tilde{r}=0$.}
\label{tab:BSd.orbits1}
\end{center}
\end{table}
\clearpage
\subsection{Analytical solution of geodesic equations}
In this subsection, we introduce the analytical solution of the equations of motion for Eqs.~(\ref{drdla'}) and (\ref{J0'}). Each equation will be discussed separately.
\subsubsection{\textbf{r motion}}
We introduce a new variable $u=\frac{1}{\tilde{r}}$, and obtain from Eq.~(\ref{drdla'}):
\begin{align}\label{15}
(\frac{du}{d\gamma})^{2}=2\tilde{L}^{2}u^{3}-\tilde{L}^{2}u^{2}+(2J^{2}+2\varepsilon)u+(\tilde{L}^{2}\tilde{\Lambda} +E^{2}-J^{2}-\varepsilon)+(\tilde{\Lambda} J^{2}+\tilde{\Lambda} \varepsilon)\dfrac{1}{u^{2}}.
\end{align}
For bouth test particles and light rays, Eq.~(\ref{15}) should be rewritten as
\begin{align}\label{16}
(u\dfrac{du}{d\gamma})^{2}=2\tilde{L}^{2}u^{5}-\tilde{L}^{2}u^{4}+(2J^{2}+2\varepsilon)u^{3}+(\tilde{L}^{2}\tilde{\Lambda} +E^{2}-J^{2}-\varepsilon)u^{2} \nonumber\\ +(\tilde{\Lambda} J^{2}+\tilde{\Lambda} \varepsilon)=\sum_{i=0}^{5}a_{i}u^{i}=\tilde{R}(u).
\end{align}
The Eq.~(\ref{16}), is hyperelliptic type and solve as follows \cite{Hackmann:2008zz,Enolski:2010if}
\begin{align}\label{17}
u(\gamma)=-\dfrac{\sigma_{1}}{\sigma_{2}}(\gamma_{\sigma}),
\end{align}
where, the argument $\gamma_{\sigma}$ is an element of the one-dimensional sigma divisor: $\gamma_{\sigma}=(f(\gamma -\gamma_{in}),\gamma -\gamma_{in})^{t}$, in which, $\gamma_{in}=\gamma_{0}+\int_{u_{0}}^{\infty}\dfrac{udu}{\sqrt{\tilde{R}(u)}}$ with $u_{0}=\dfrac{1}{\tilde{r}_{0}}$ depends only on the initial values, and the function $f$ is given by the condition $\sigma(\gamma_{\sigma})=0$. Also, $\sigma_{i}$, is the $i$-th derivative of the Kleinian sigma function in two variables
\begin{align}
\sigma(z)=C e^{z^{t}k z}\theta[K_{\infty}](2\omega^{-1}z;\tau),
\end{align}
which is given by the Riemann $\theta$-function with characteristic $K_{\infty}$, in which $k=\eta(2\omega)^{-1}$, $(2\omega, 2\omega')$, is the period-matrix, $(2\eta,2\eta')$, is the period-matrix of the second kind, $\tau$, is  the symmetric Riemann matrix, $C$, is the constant and $2K_{\infty}=(0,1)^{t}+(1,1)^{t}\tau$, is the vector of Riemann constants with base point at infinity.
For more details on the construction of such solutions see e.g.\cite{Enolski:2010if,Buchstaber:2012wb}. 
Finally, the solution for $\tilde{r}$ with use of Eq.~(\ref{17}), is given by
\begin{align}
\tilde{r}=-\dfrac{\sigma_{2}}{\sigma_{1}}(\gamma_{\sigma}).
\end{align}
\subsubsection{\textbf{w motion}}
We substitute $d\gamma=\dfrac{u du}{\sqrt{\tilde{R}(u)}}$ and $r=\frac{1}{u}$, in Eqs.~(\ref{J0'}) and (\ref{16}) and obtain
\begin{align}
\tilde{w}-\tilde{w}_{0}=J\int_{u_{0}}^{u}\dfrac{du}{u\sqrt{\tilde{R}(u)}}.
\end{align}
This integral can be expressed in terms of the canonical integral of third kind $\int dP(x_{1},x_{2})$ \cite{Hackmann:2008zz,Hackmann:2010zz}, In particular, we have
\begin{align}
\int_{u_{0}}^{u}\dfrac{du}{(u-u_{i})\sqrt{\tilde{R}(u)}}=\dfrac{1}{+\sqrt{\tilde{R}(u_{i})}}\int_{u_{0}}^{u}dP(u_{i}^{+},u_{i}^{-}).
\end{align}
So, the solution for $ w $ is 
\begin{align}
w=\dfrac{J}{\sqrt{\tilde{R}(u_{i})}}\bigg[\frac{1}{2}\log\dfrac{\sigma(W^{+}(\omega))}{\sigma(W^{-}(\omega))}-\frac{1}{2}\log\dfrac{\sigma(W^{+}(\omega_{0}))}{\sigma(W^{-}(\omega_{0}))}\nonumber\\ -(f(\omega)-f(\omega_{0}), \omega - \omega_{0}) \big( \int_{u_{i}^{-}}^{u_{i}^{+}} d\vec{r} \big) \bigg] +w_{0}
\end{align}
\subsection{Orbits}
\begin{figure}[!ht]
\centering

\subfigure[]{
\includegraphics[width=6cm]{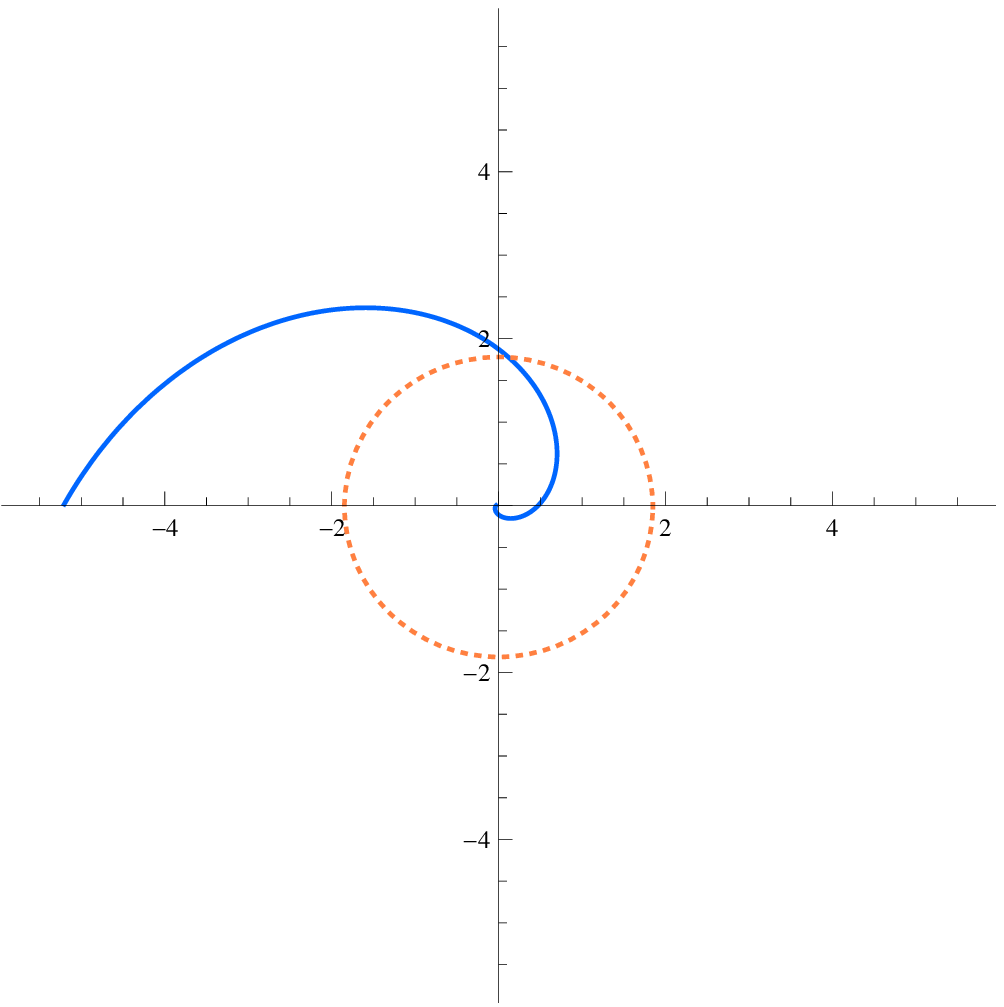}
	\label{23}
}
\subfigure[]{
\includegraphics[width=6cm]{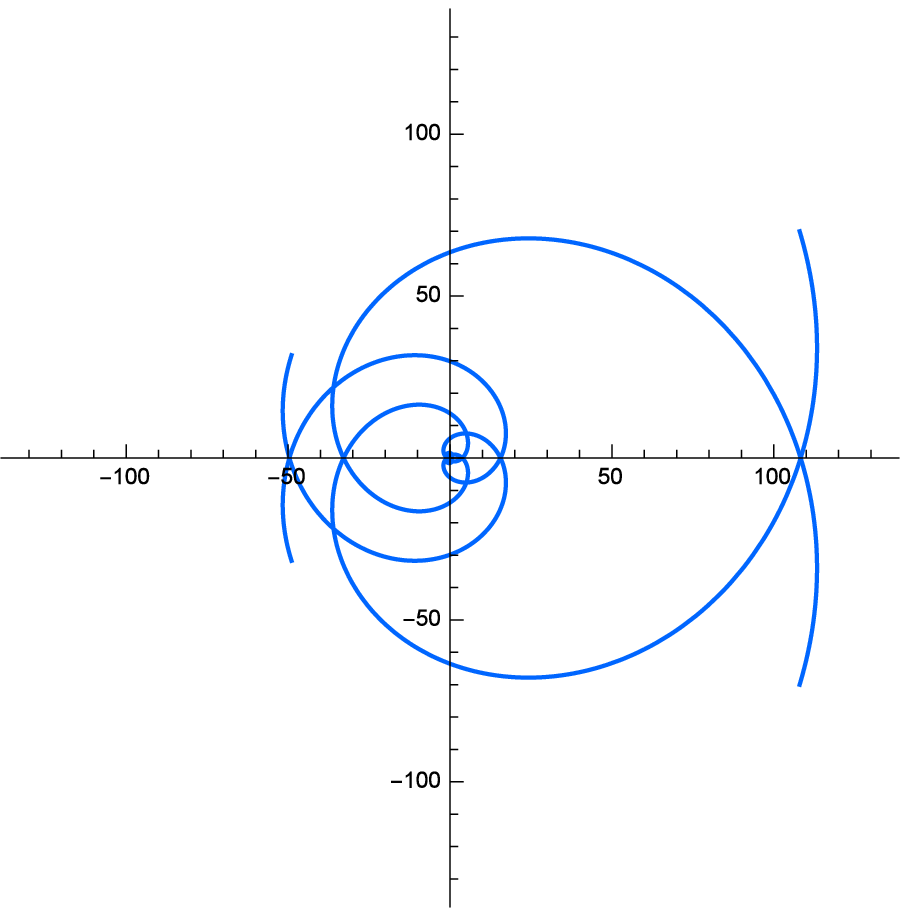}
	\label{24}
}
\subfigure[]{
\includegraphics[width=6cm]{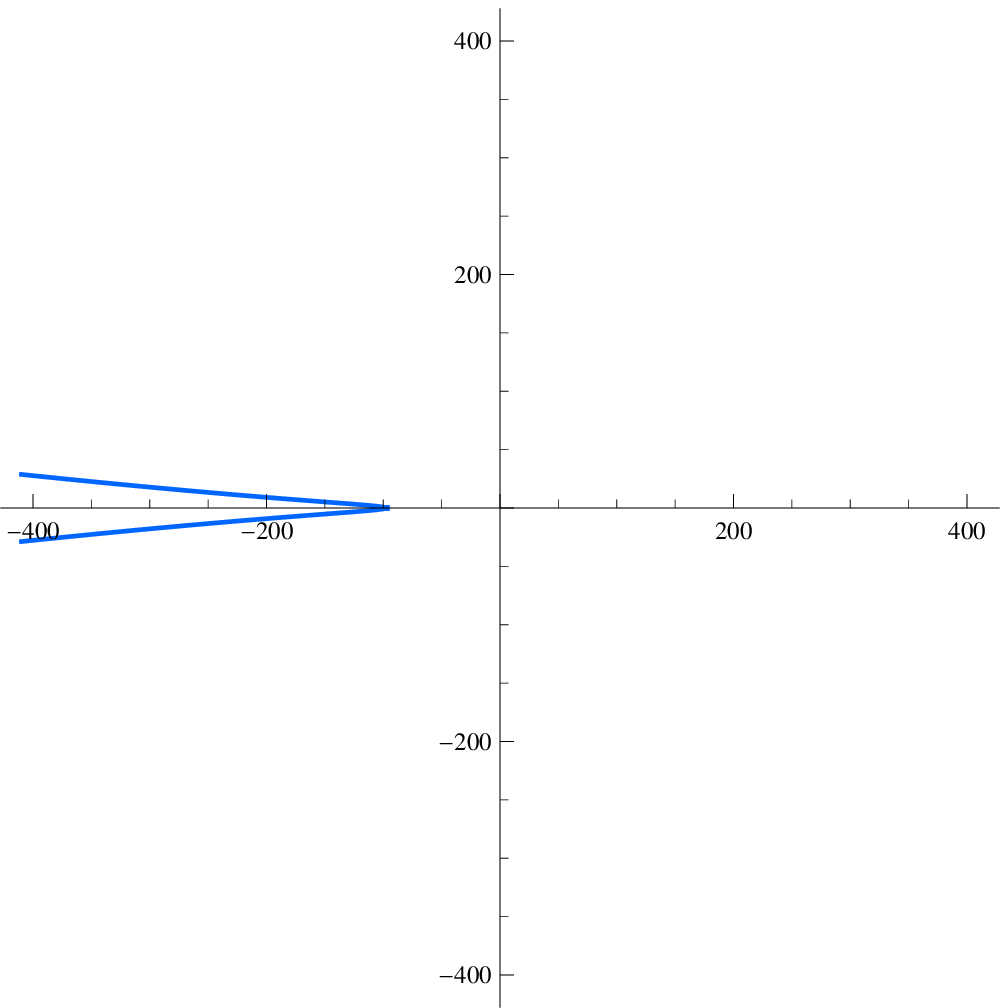}
	\label{25}
}
\caption[figs]
{Different types of orbits in black string (anti)-de sitter spacetime: 
\subref{23}, terminating Orbit (TO) in Region II for, $E=\sqrt{1.5}$, $L=\frac{1}{9}$,  $\varepsilon=1$, $\Lambda=\frac{1}{3}\times 10^{-5}$, $J=1$.
\subref{24}, Bound orbits (BO) in Region III for, $\varepsilon=1$, $L=5.8$, $E=\sqrt{1.9}$, $\Lambda=\frac{1}{3}\times 10^{-5}$, $J=1$.
\subref{25}, Escape orbits (EO) in Region III for, $\varepsilon=1$, $L=5.8$, $E=\sqrt{1.9}$, $\Lambda=\frac{1}{3}\times 10^{-5}$, $J=1$.} 
\label{26}
\end{figure}

With these analytical results, with the help of parametric $ \tilde{L}-E^{2} $-diagrams, (Figs.~\ref{3} and \ref{6}), and effective potential diagram, (Figs.\ref{11} and \ref{19}), we plot same example of possible orbit types in the static black string-(anti-) de sitter spacetime, which are shown in Fig.~\ref{26}.
It can be seen from Fig.~\ref{26}(a), we have TO motion in region I. Also, example of BO, motion is presented in Fig.~\ref{26}(b), for region  III. Moreover, example of EO, motion can be observed in Fig.~\ref{26}(c).

\clearpage

\section{rotating black string-(anti-) de sitter spacetime}
In this section, we study the geodesics in the rotating black string-(anti-) de sitter spacetime and introduce analytical solutions of the equations of motion and orbits types. We add an extra compact dimension to the Kerr-(anti-) de sitter metric then derive the rotating black string-(anti-) de sitter metric. The whole solution of the geodesics in the Kerr-(anti-) de sitter spacetime can be found in Ref.~\cite{Hackmann:2010zz}.
\subsection{The geodesic equations}
If we add an extra compact spatial dimension $w$, to the kerr-(anti-) de sitter metric, then the metric takes the form
\begin{align}\label{RBS1}
ds^{2}=-\dfrac{\Delta_{r}}{\chi^{2}\rho^{2}} \big( dt -a \sin^{2}\theta d\varphi \big)^{2} +\dfrac{\rho^{2}}{\Delta_{r}}dr^{2}+\dfrac{\Delta_{\theta}\sin^{2}\theta}{\chi^{2}\rho^{2}} \big(a dt -(r^{2}+a^{2}) d\varphi \big)^{2} +\dfrac{\rho^{2}}{\Delta_{\theta}}d\theta^{2}+ dw^{2}.
\end{align}
where
\begin{align}
\Delta_{r}= \big(1-\frac{\Lambda}{3}r^{2} \big) (r^{2}+a^{2})- 2Mr, \qquad \Delta_{\theta}=1+\frac{a^{2}\Lambda}{3}\cos^{2}\theta , \qquad \nonumber\\ \chi =1+\frac{a^{2}\Lambda}{3}, \qquad  \rho^{2}=r^{2}+a^{2} \cos^{2}\theta .
\end{align}
This solution describes a rotating black string-(anti-) de sitter spacetime. $M$ is proportional to the mass of the black string-(anti-) de sitter, $a$ is proportional to the angular momentum, and $\Lambda$ is the cosmological constant. Notice that this metric has coordinate singularities on the axes $\theta=0,\pi$ and there are two
horizons defined by $\Delta_{r}=0$. With assuming that $a\neq0$, the only real singularity is located at $\rho^{2}=0$, i.e. at simultaneously $r=0$ and $\theta=\frac{\pi}{2}$.
\\
The Hamilton-Jacobi equation 
\begin{equation}\label{Hamilton}
\dfrac{\partial S}{\partial\tau}+\frac{1}{2}\ g^{\mu\nu}\dfrac{\partial S}{\partial X^{\mu}}\dfrac{\partial S}{\partial X^{\nu}}=0.
\end{equation}
can be solved with and ansatz for the action
\begin{equation}
\label{S}
S=\frac{1}{2}\varepsilon \tau - Et+L\varphi +Jw+S_{\theta}(\theta) + S_{r} (r) ,
\end{equation}
where $\tau$ is an affine parameter along the geodesic. The parameter $\varepsilon$, is equal to one for particles and equal to zero for light.

using Eqs.~(\ref{RBS1})--(\ref{Hamilton}), we have

\begin{align}\label{ds/dr.ds/dtheta}
\varepsilon a^{2} \cos^{2}\theta +J^{2}a^{2}\cos^{2}\theta +\Delta_{\theta}(\dfrac{ds}{d\theta})^{2}+\dfrac{\chi^{2}}{\Delta_{\theta}\sin^{2}\theta} \big( aE \sin^{2}\theta - L\big)^{2}= \nonumber\\ -\varepsilon r^{2}-r^{2}J^{2}-\Delta_{r}(\dfrac{ds}{dr})^{2}+\dfrac{\chi^{2}}{\Delta_{r}} \big( (r^{2}+a^{2})E-aL \big)^{2}  ,
\end{align}
 Each side of this equation~(\ref{ds/dr.ds/dtheta}), only depends on $r$ or $\theta$. This means that each side is equal to the famous Carter constant $K$ \cite{Carter:1968rr}. \\
From the separation ansatz Eq.~(\ref{S}), we derive the equations of motion
\begin{align}\label{Joda}
\rho^{4}(\dfrac{dr}{d\tau})^{2}=\chi^{2}\big((r^{2}+a^{2})E -aL\big)^{2}-\Delta_{r} \big( K+(\varepsilon +J^{2})r^{2} \big)=R(r) ,
\end{align}
\begin{align}\label{Jodatheta}
\rho^{4}(\dfrac{d\theta}{d\tau})^{2}=\Delta_{\theta} \big( K-\varepsilon a^{2}\cos^{2}\theta -J^{2}a^{2}\cos^{2}\theta \big) -\frac{\chi^{2}}{\sin^{2}\theta} \big( aE \sin^{2}\theta -L \big)^{2} =\Theta(\theta),
\end{align}
\begin{align}
\dfrac{\rho^{2}}{\chi^{2}}(\dfrac{d\varphi}{d\tau})=\dfrac{E(a^{3}+r^{2}a)-L a^{2}}{\Delta_{r}}+\dfrac{1}{\Delta_{\theta}\sin^{2}\theta} \big( L-aE \sin^{2}\theta \big) ,
\end{align}
\begin{align}\label{dt}
\dfrac{\rho^{2}}{\chi^{2}}(\dfrac{dt}{d\tau})=\dfrac{L(-a^{3}-r^{2}a)+E(a^{2}+r^{2})^{2}}{\Delta_{r}}+\dfrac{(aL-E a^{2}\sin^{2}\theta)}{\Delta_{\theta}},
\end{align}
\begin{align}\label{J}
\rho^{2}(\dfrac{dw}{d\tau})=\rho^{2}J.
\end{align}
 In the following, we will explicitly solve these equations. Eq.~(\ref{Joda}) suggests the introduction of an effective potential $V_{eff,r}$, such that $V_{eff,r}=E$, corresponds to $(\dfrac{dr}{d\tau})^{2}=0$. However, in contrast to the spherically symmetric case, there are two solutions 
\begin{equation}
V_{eff,r}^{\pm}=\dfrac{\chi L a \pm \sqrt{\Delta_{r}(K+\varepsilon r^{2} + J^{2}r^{2})}}{(a^{2}+r^{2})\chi},
\end{equation} 
 where $(\dfrac{dr}{d\tau})^{2}\geq 0$ for $E\leq V_{eff,r}^{-}$ and $E\geq V_{eff,r}^{+}$. In the same way an effective potential
corresponding to Eq.~(\ref{Jodatheta}) can be introduced
\begin{align}
V_{eff,\theta}^{\pm}=\dfrac{L\chi \pm \sqrt{\Delta_{\theta}\sin^{2}\theta (K-\varepsilon a^{2}\cos^{2}\theta -J^{2}a^{2}\cos^{2}\theta)}}{a \chi \sin^{2}\theta}.
\end{align}
but here, $(\dfrac{d\theta}{d\tau})^{2}\geq 0$ for $V_{eff,\theta}^{-}\leq E \leq V_{eff,\theta}^{+}$.\\
The geodesic equations Eq.~(\ref{Joda})--(\ref{J}), are coupled by $\rho^{2}=r^{2}+a^{2}\cos^{2}\theta$. Introducing the Mino time $\lambda$ \cite{Mino:2003yg} connected to the proper time $\tau$ by $\dfrac{d\tau}{d\lambda}=\rho^{2}$, the equations of motions take the forms
\begin{align}\label{d}
(\dfrac{dr}{d\lambda})^{2}=R(r)=\chi^{2}\big((r^{2}+a^{2})E -aL\big)^{2}-\Delta_{r} \big( K+(\varepsilon +J^{2})r^{2} \big),
\end{align}
\begin{align}
(\dfrac{d\theta}{d\lambda})^{2}=\Theta(\theta)=\Delta_{\theta} \big( K-\varepsilon a^{2}\cos^{2}\theta -J^{2}a^{2}\cos^{2}\theta \big) -\frac{\chi^{2}}{\sin^{2}\theta} \big( aE \sin^{2}\theta -L \big)^{2},
\end{align}
\begin{align}
\dfrac{1}{\chi^{2}}(\dfrac{d\varphi}{d\lambda})=\dfrac{E(a^{3}+r^{2}a)-L a^{2}}{\Delta_{r}}+\dfrac{1}{\Delta_{\theta}\sin^{2}\theta} \big( L-aE \sin^{2}\theta \big),
\end{align}
\begin{align}\label{t}
\dfrac{1}{\chi^{2}}(\dfrac{dt}{d\lambda})=\dfrac{L(-a^{3}-r^{2}a)+E(a^{2}+r^{2})^{2}}{\Delta_{r}}+\dfrac{(aL-E a^{2}\sin^{2}\theta)}{\Delta_{\theta}},
\end{align}
\begin{align}\label{JJ}
(\dfrac{dw}{d\lambda})=\rho^{2}J.
\end{align}
Again for simplicity, we rescale the parameters appearing in Eqs.~(\ref{d})--(\ref{JJ}), wich dimensionless parameters
\begin{align}
\tilde{r}=\dfrac{r}{M} , \qquad \tilde{a}=\dfrac{a}{M} , \qquad \tilde{t}=\dfrac{t}{M} , \qquad \tilde{L}=\dfrac{L}{M},\qquad \tilde{\Lambda}=\frac{1}{3}\Lambda M^{2}, \nonumber\\ \tilde{w}=\dfrac{w}{M},\qquad \tilde{K}=\dfrac{K}{M^{2}}, \qquad \gamma = M\lambda ,
\end{align}
and accordingly
\begin{align}
\Delta_{\tilde{r}}=(1-\tilde{\Lambda}\tilde{r}^{2})(\tilde{r}^{2}+\tilde{a}^{2})-2\tilde{r}, \qquad \Delta_{r}=M^{2}\Delta_{\tilde{r}},  \qquad \Delta_{\theta}=1+\tilde{a}^{2}\tilde{\Lambda}\cos^{2}\theta , \nonumber\\ \tilde{\rho}^{2}=\tilde{a}^{2}+\tilde{r}^{2}\cos^{2}\theta , \qquad \chi =1+\tilde{a}^{2}\tilde{\Lambda}.
\end{align}
Then, the equations~(\ref{d})--(\ref{JJ}) can be rewritten as
\begin{align}\label{drd}
(\dfrac{d\tilde{r}}{d\gamma})^{2}=\chi^{2}P^{2}(r)-\Delta_{\tilde{r}}\big(\varepsilon \tilde{r}^{2}+\tilde{K}+J^{2}\tilde{r}^{2} \big) =\tilde{R}(\tilde{r}) ,
\end{align}
\begin{align}\label{dthetad}
(\dfrac{d\theta}{d\gamma})^{2}=\Delta_{\theta} \big( \tilde{K}-\varepsilon \tilde{a}^{2}\cos^{2}\theta -J^{2}\tilde{a}^{2}\cos^{2}\theta \big)-\dfrac{\chi^{2}T^2(\theta)}{\sin^{2}\theta}=\tilde{\Theta}(\theta),
\end{align}
\begin{align}\label{dphi}
\dfrac{1}{\chi^{2}}(\dfrac{d\varphi}{d\gamma})=\dfrac{\tilde{a}}{\Delta_{\tilde{r}}}P(r)-\dfrac{1}{\Delta_{\tilde{\theta}}\sin^{2}\theta}T(\theta) ,
\end{align}
\begin{align}\label{dtd}
\dfrac{1}{\chi^{2}}(\dfrac{d\tilde{t}}{d\gamma})=\dfrac{\tilde{r}^{2}+\tilde{a}^{2}}{\Delta_{\tilde{r}}}P(r)-\dfrac{\tilde{a} }{\Delta_{\tilde{\theta}}}T(\theta) ,
\end{align}
\begin{align}\label{JJJ}
(\dfrac{d\tilde{w}}{d\gamma})=\tilde{\rho}^{2}J,
\end{align}
where
\begin{align*}P(r)=(\tilde{r}^{2}+\tilde{a}^{2})E-\tilde{a}\tilde{L}, \nonumber\\ T(\theta)=\tilde{a}E \sin^{2}\theta -\tilde{L} .
\end{align*}
In section \ref{ana}, we will explicitly solve these equations.
\begin{align}\label{VR}
V_{eff,r}^{\pm}=\dfrac{\tilde{L}\tilde{\Lambda}\tilde{a}^{3}+\tilde{a}\tilde{L}\pm\sqrt{(\tilde{\Lambda}\tilde{a}^{2}\tilde{r}^{2}+\tilde{\Lambda}\tilde{r}^{4}-\tilde{a}^{2}-\tilde{r}^{2}+2\tilde{r})(-J^{2}\tilde{r}^{2}-\varepsilon\tilde{r}^{2}-\tilde{K})}}{(\tilde{a}^{2}+\tilde{r}^{2})(\tilde{\Lambda}\tilde{a}^{2}+1)}
\end{align}
\begin{align}
V_{eff,\theta}^{\pm}=\dfrac{\tilde{L}\chi \pm\sqrt{\Delta_{\theta}(-J^{2}\tilde{a}^{2}\cos^{2}\theta -\varepsilon\tilde{a}^{2}\cos^{2}\theta +\tilde{K})\sin^{2}\theta}}{\tilde{a}\chi \sin^{2}\theta}
\end{align}

\subsection{Types of latitudinal motion}
 
First we substitute $\nu =\cos^{2}\theta$  in the function $\tilde{\Theta}(\theta)$:
 \begin{align}\label{thethaL}
 \tilde{\Theta}(\nu)=(1+ \tilde{a}^{2}\tilde{\Lambda}\nu)(\tilde{K}-\varepsilon \tilde{a}^{2} \nu -J^{2}\tilde{a}^{2}\nu)- \chi^{2} \big(\tilde{a}^{2}E^{2}(1-\nu)-2\tilde{L}\tilde{a}E+\dfrac{\tilde{L}^{2}}{(1-\nu)} \big),
 \end{align}
In order to specify the number of real zeros of $\tilde{\Theta}(\nu)$ in $[0, 1]$, we suppose that for a given set of parameters, there exists in [0, 1] a certain number of zeros for $\tilde{\Theta}(\nu)$. $\nu=0$ is a zero of $\tilde{\Theta}$ if
 \begin{align}
 \tilde{\Theta}(\nu =0)=\tilde{K}- \chi^{2}(\tilde{a}E-\tilde{L})^{2}=0,
 \end{align}
 and therefore
 \begin{align}\label{Ly}
 \tilde{L}=\tilde{a}E \pm \dfrac{\sqrt{\tilde{K}}}{\chi},
 \end{align}
 As $\nu=1$ is a pole of $\tilde{\Theta}(\nu)$ for $\tilde{L}\neq 0$, it is only possible that $\nu=1$ is a zero of $\tilde{\Theta}(\nu)$ if $\tilde{L}=0$,
 \begin{align}
 \tilde{\Theta}(\nu =1, \tilde{L}=0)=(1+\tilde{a}^{2}\tilde{\Lambda})(\tilde{K}-\varepsilon \tilde{a}^{2}-J^{2}\tilde{a}^{2})=\chi (\tilde{K}-J^{2}\tilde{a}^{2}-\varepsilon \tilde{a}^{2}),
 \end{align}
 To remove the pole of $\tilde{\Theta}(\nu)$ at $\nu=1$ we consider
 \begin{align}\label{thetanoo}
 \tilde{\Theta'}_{\nu}=(1-\nu)(1+\tilde{a}^{2}\tilde{\Lambda}\nu)(\tilde{K}-\varepsilon \tilde{a}^{2}\upsilon -J^{2}\tilde{a}^{2}\nu) - \chi^{2}(\tilde{a}E(1-\nu)-\tilde{L})^{2} ,
 \end{align}
here $\tilde{\Theta}(\nu)=\frac{1}{1-\nu}\tilde{\Theta'}(\nu)$. Then double zeros fulfil the conditions
 \begin{align}\label{x}
 \tilde{\Theta'}_{\nu}=0, \qquad \dfrac{d\tilde{\Theta'}_{\nu}}{d\nu}=0,
 \end{align}
 which in lead to
 \begin{align}\label{LLL}
 \tilde{L}=\dfrac{E \chi \pm \sqrt{\tilde{K}\tilde{\Lambda}+\chi^{2}E^{2}-J^{2}\tilde{\Lambda}^{2}\tilde{a}^{2}}}{2 \tilde{a}\tilde{\Lambda}}.
 \end{align}

With the help of Eqs.~(\ref{Ly})--(\ref{LLL}), parametric $\tilde{L}-E^{2}$ -diagrams can be drawn (see Fig.~\ref{27}). Below we give a list of possible regions:
\begin{figure}[!ht]
\centering
\includegraphics[width=8cm]{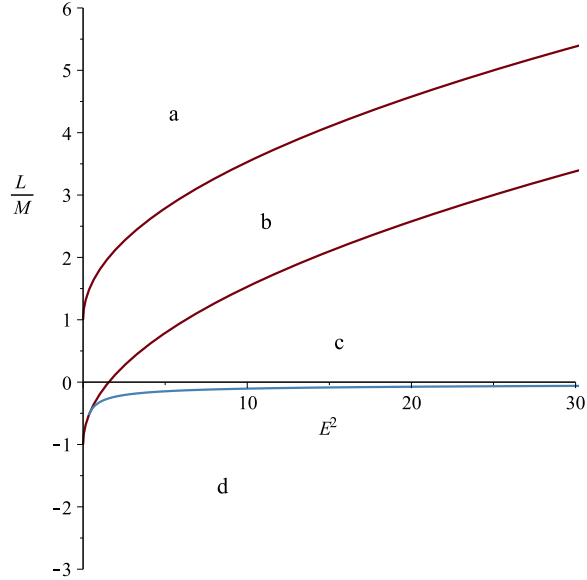}
	
\caption[figs]
{ Parametric $\tilde{L}-E^{2}$ -diagrams for $\theta$-motion with, $\varepsilon=1$, $a=0.8$, $J=1$, $\Lambda=10^{-5}$.}
\label{27}
\end{figure}

\begin{enumerate}
\item[a]:~no geodesic motion possible
\item[b]:~$\tilde{\Theta}_{\nu}$ has one real zero $\nu_{max}$ in $[0, 1)$ with $\tilde{\Theta}_{\nu}\geqslant 0$ for $\nu\in[0,\nu_{max}]$, i.e. $\theta$ oscillates
around the equatorial plane $\theta=\frac{\pi}{2}$
\item[c]:~$\tilde{\Theta}_{\nu}$ has two real zero $\nu_{min}$, $\nu_{max}$ in $[0, 1)$ with $\tilde{\Theta}_{\nu}\geqslant 0$ for $\nu\in[\nu_{min},\nu_{max}]$, i.e. $\theta$ oscillates between $\arccos(\pm\sqrt{\nu_{min}})$ and $\arccos(\pm\sqrt{\nu_{max}})$.
\item[d]:~no geodesic motion possible,
\end{enumerate}

\subsection{Types of radial motion}
\begin{figure}[!ht]
\centering
\subfigure[]{
\includegraphics[width=6cm]{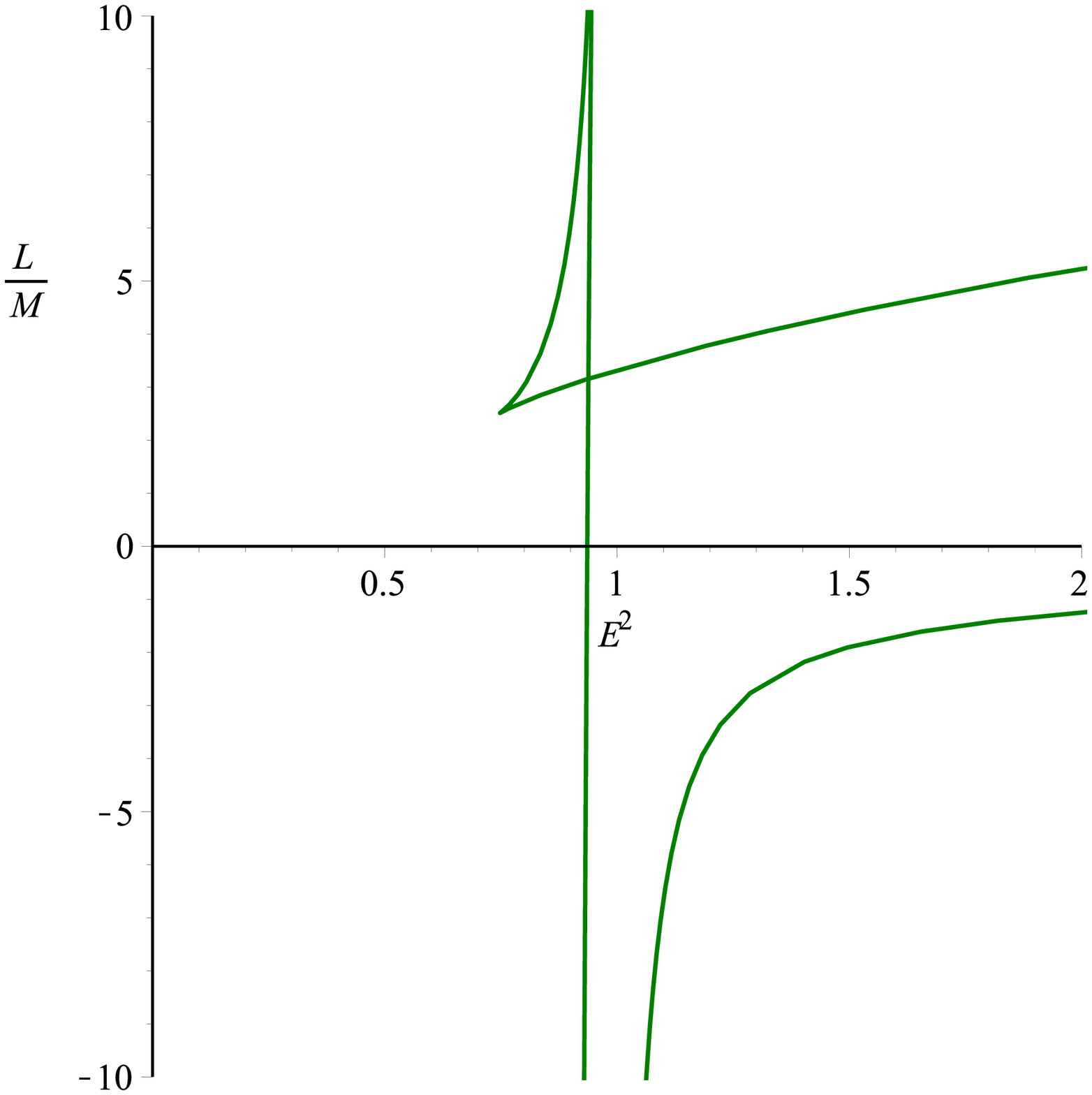}
	\label{28}
}
\hspace*{20mm}
\subfigure[]{
\includegraphics[width=6cm]{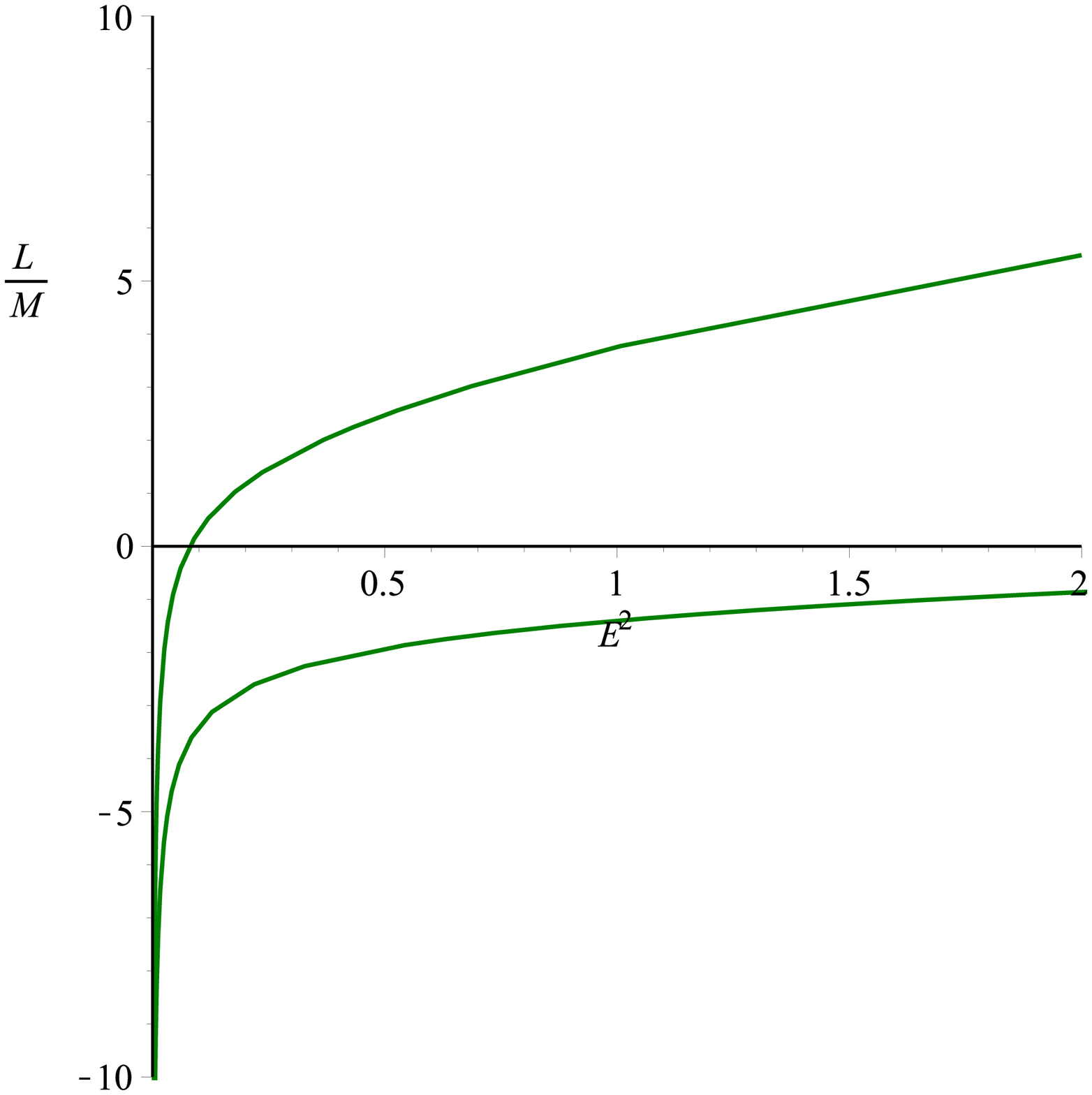}
	\label{29}
}

\subfigure[]{
\includegraphics[width=6cm]{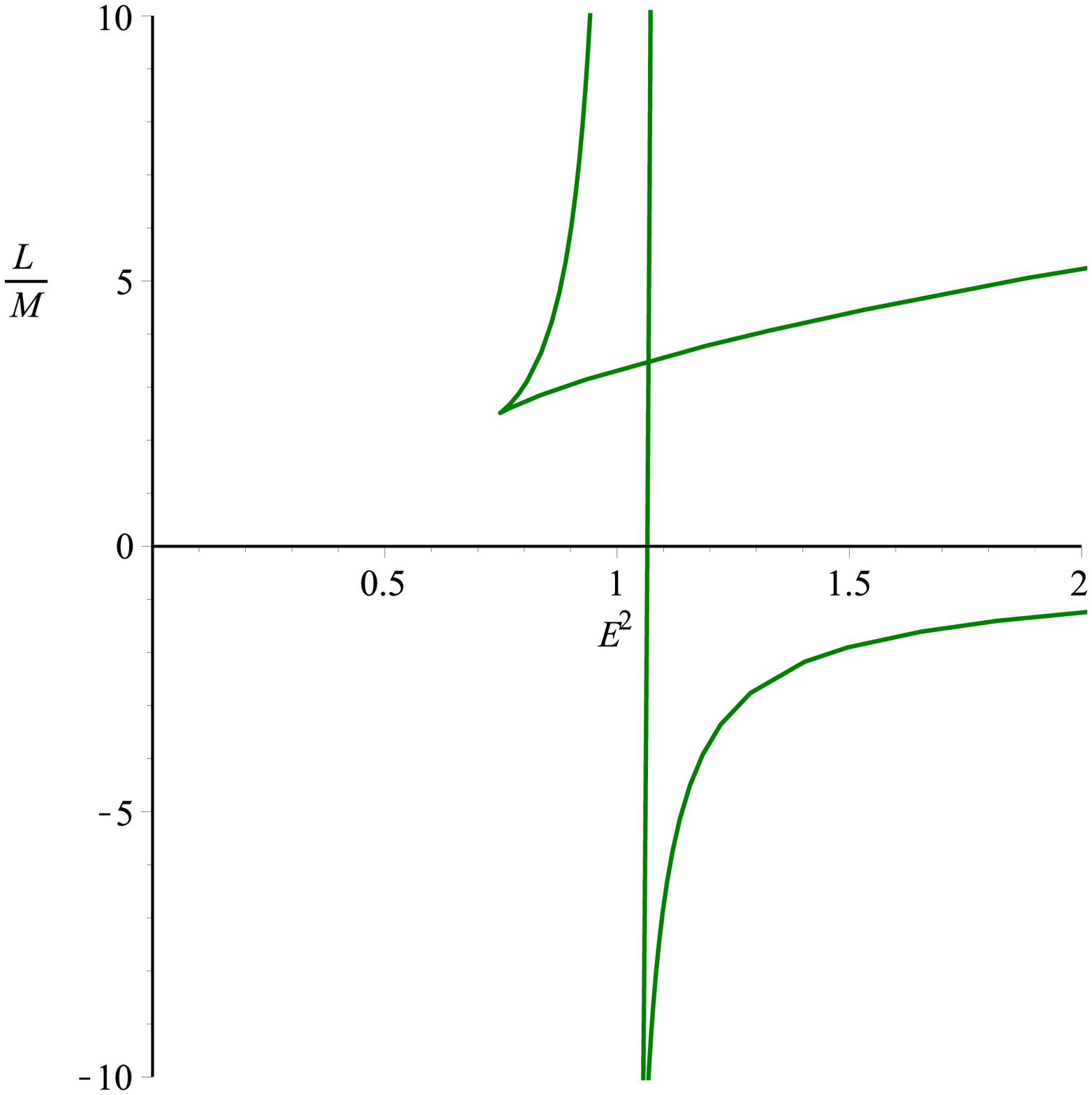}
	\label{30}
}
\hspace*{20mm}
\subfigure[]{
\includegraphics[width=6cm]{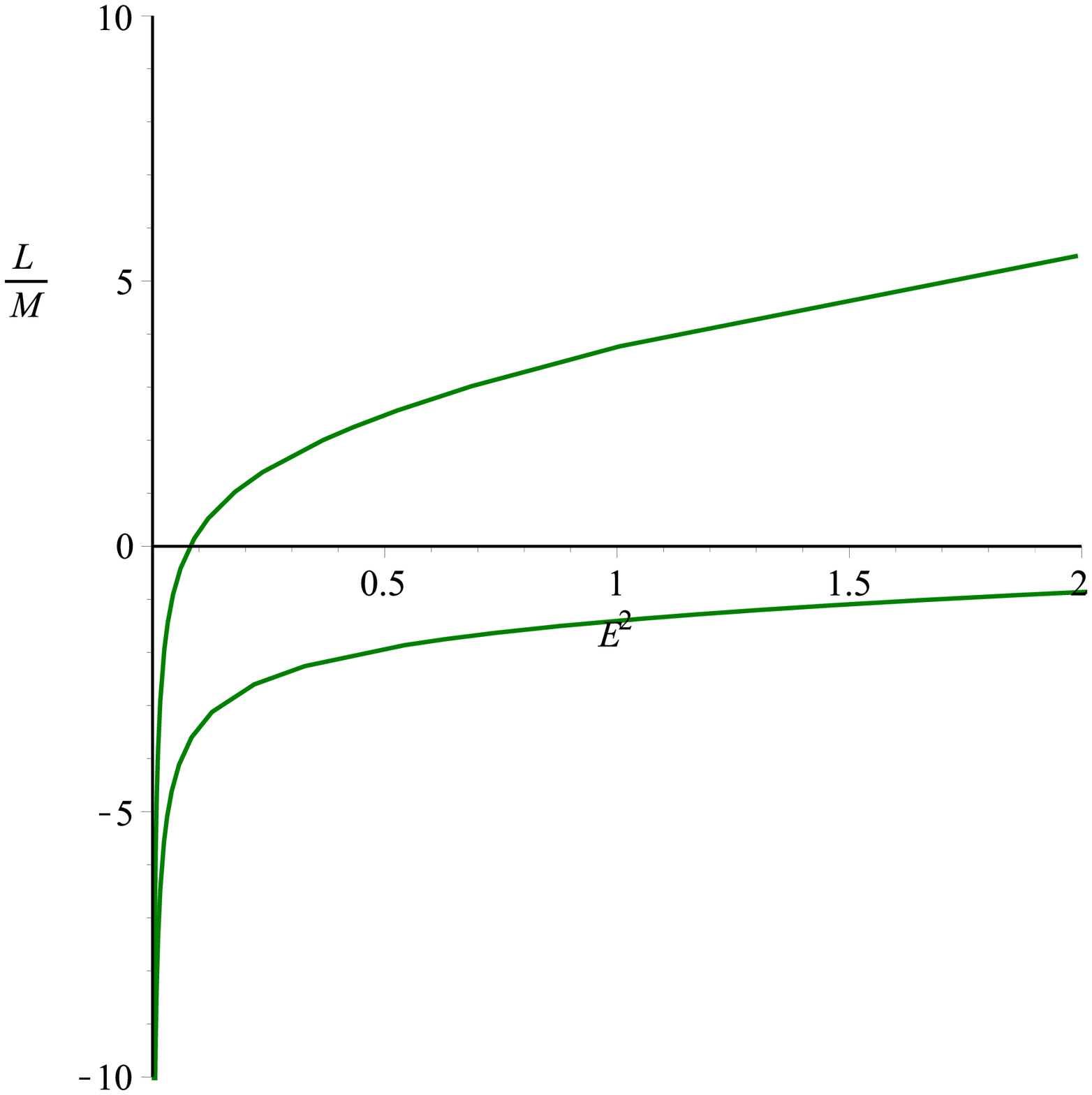}
	\label{31}
}
\caption[figs]
{Parametric $\tilde{L}-E^{2}$-diagram of the $\tilde{r}$-motion. \subref{28} kerr-de sitter, with $\tilde{a}=0.8$, $\tilde{\Lambda}=10^{-5}$, $\tilde{K}=2$, $J=0$. 
 \subref{29} rotating black string-de sitter, with $\tilde{a}=0.8$, $\tilde{\Lambda}=10^{-5}$, $\tilde{K}=2$, $J=1$.
  \subref{30} kerr-anti de sitter, with $\tilde{a}=0.8$, $\tilde{\Lambda}=-10^{-5}$, $\tilde{K}=2$, $J=0$.
 \subref{31} rotating black string-anti de sitter, with $\tilde{a}=0.8$, $\tilde{\Lambda}=-10^{-5}$, $\tilde{K}=2$, $J=1$.} 
\label{32}
\end{figure}

 In this subsection, we use of $\tilde{R}(\tilde{r})$ in equation~(\ref{drd}) to determine orbit types.
\begin{align}
\tilde{R}(\tilde{r})=\chi^{2}((\tilde{r}^{2}+\tilde{a}^{2})E-\tilde{a}\tilde{L})^{2} - \Delta_{\tilde{r}}(\varepsilon \tilde{r}^{2}+\tilde{K}+J^{2}\tilde{r}^{2}).
\end{align}
The zeros of the polynomial $\tilde{R}(\tilde{r})$ are the turning points of orbits of light and test particles, the number of zeros can only change if double zeros occur, if
\begin{align}\label{dRdd}
\tilde{R}(\tilde{r})=0, \qquad \dfrac{d\tilde{R}(\tilde{r})}{d\tilde{r}}=0.
\end{align}
with the help of conditions (\ref{dRdd}), parametric $\tilde{L}-E^{2}$-diagrams can be drawn, (see Fig.\ref{32})
\newpage

\newpage
\begin{figure}[!ht]
\centering
\subfigure[]{
\includegraphics[width=7cm]{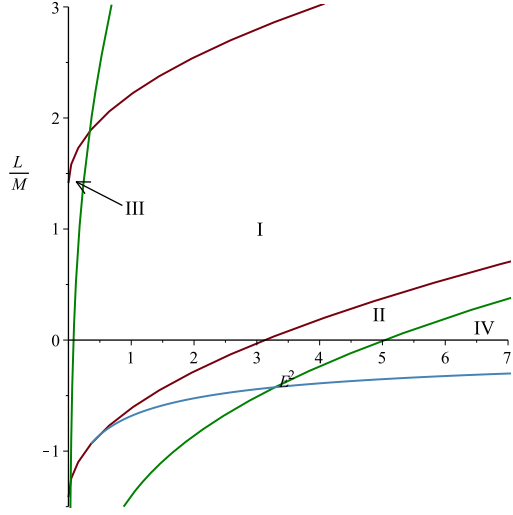}
	\label{33}
}

\subfigure[]{
\includegraphics[width=7cm]{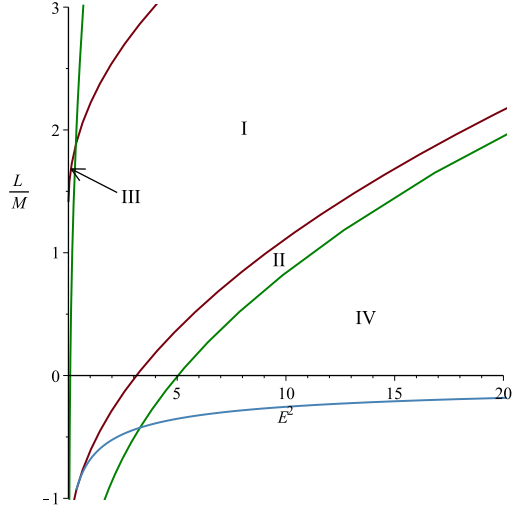}
	\label{34}
}

\caption[figs]
{Combined parametric $\tilde{L}-E^{2}$-diagram of the $\theta$-motion and the $\tilde{r}$-motion. \subref{33} $\varepsilon =1$, $\tilde{a}=0.8$, $J=1$, $\tilde{K}=2$, $\tilde{\Lambda}=10^{-5}$ . 
 \subref{34} Different view of figure a.} 
\label{35}
\end{figure}
\begin{figure}[!ht]
\centering
\subfigure[ ]{
\includegraphics[width=6.5cm]{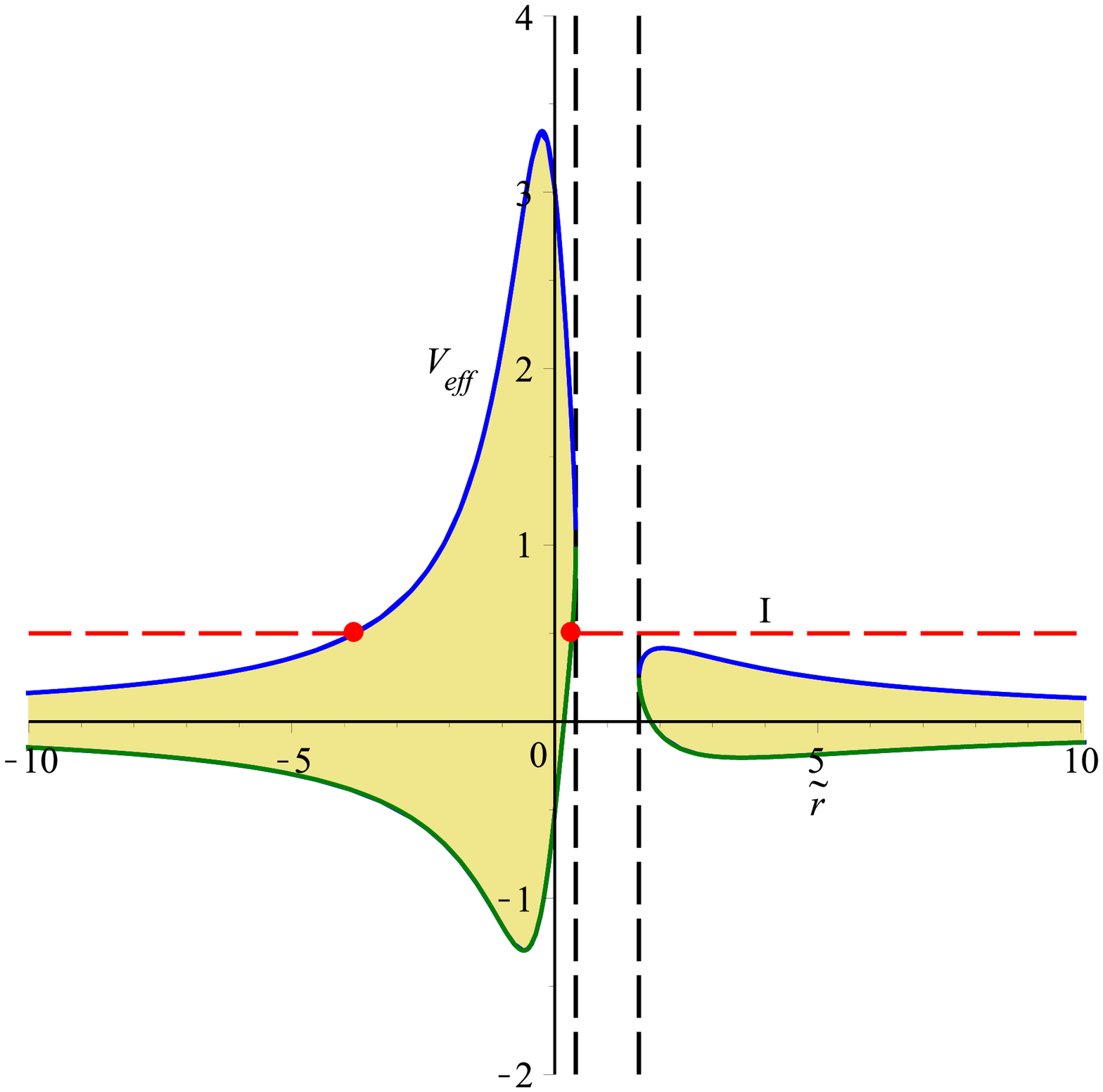}
	\label{36}
}
\hspace*{10mm}
\subfigure[ ]{
\includegraphics[width=6.5cm]{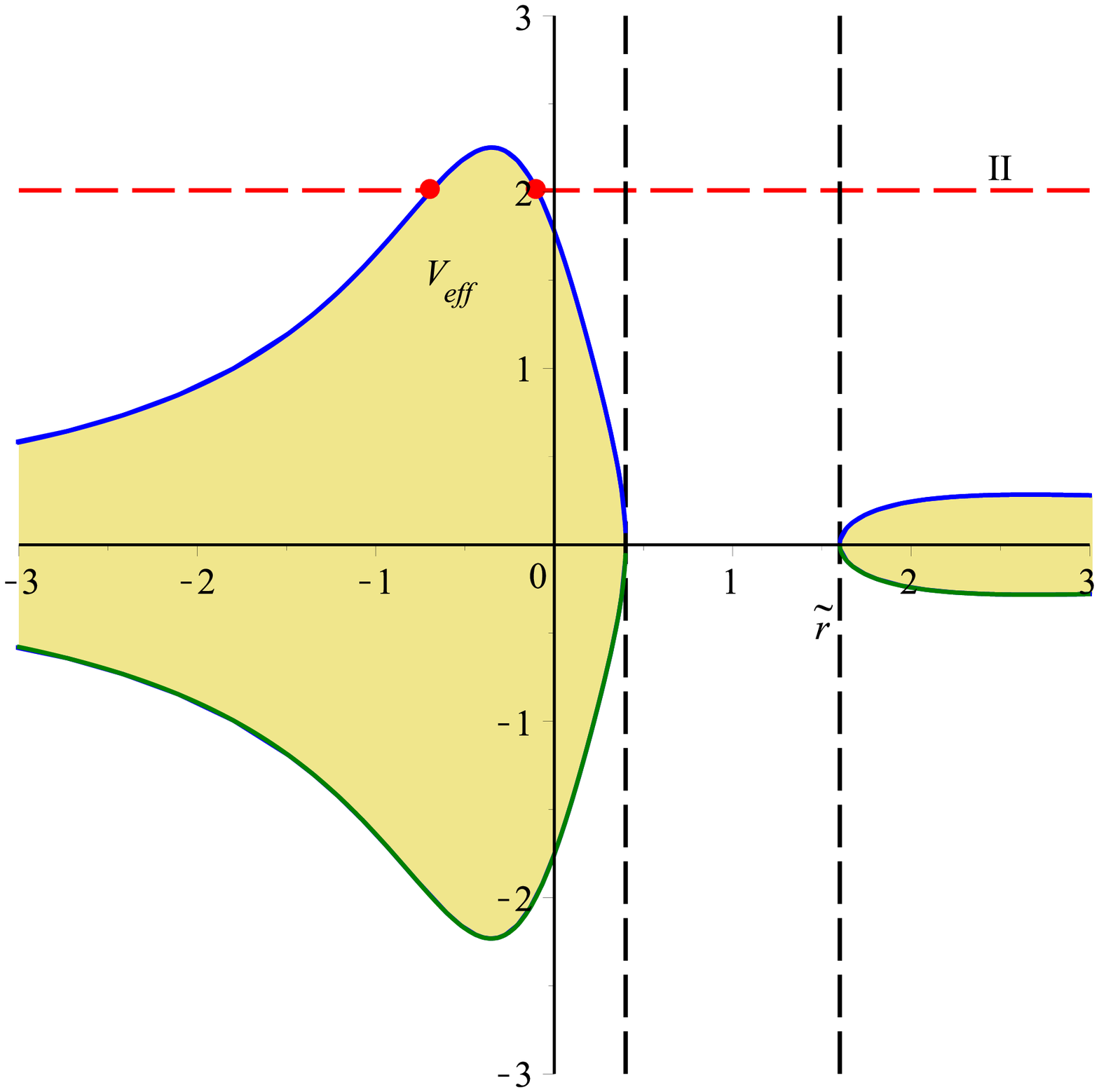}
	\label{37}
}
\subfigure[ ]{
\includegraphics[width=6.5cm]{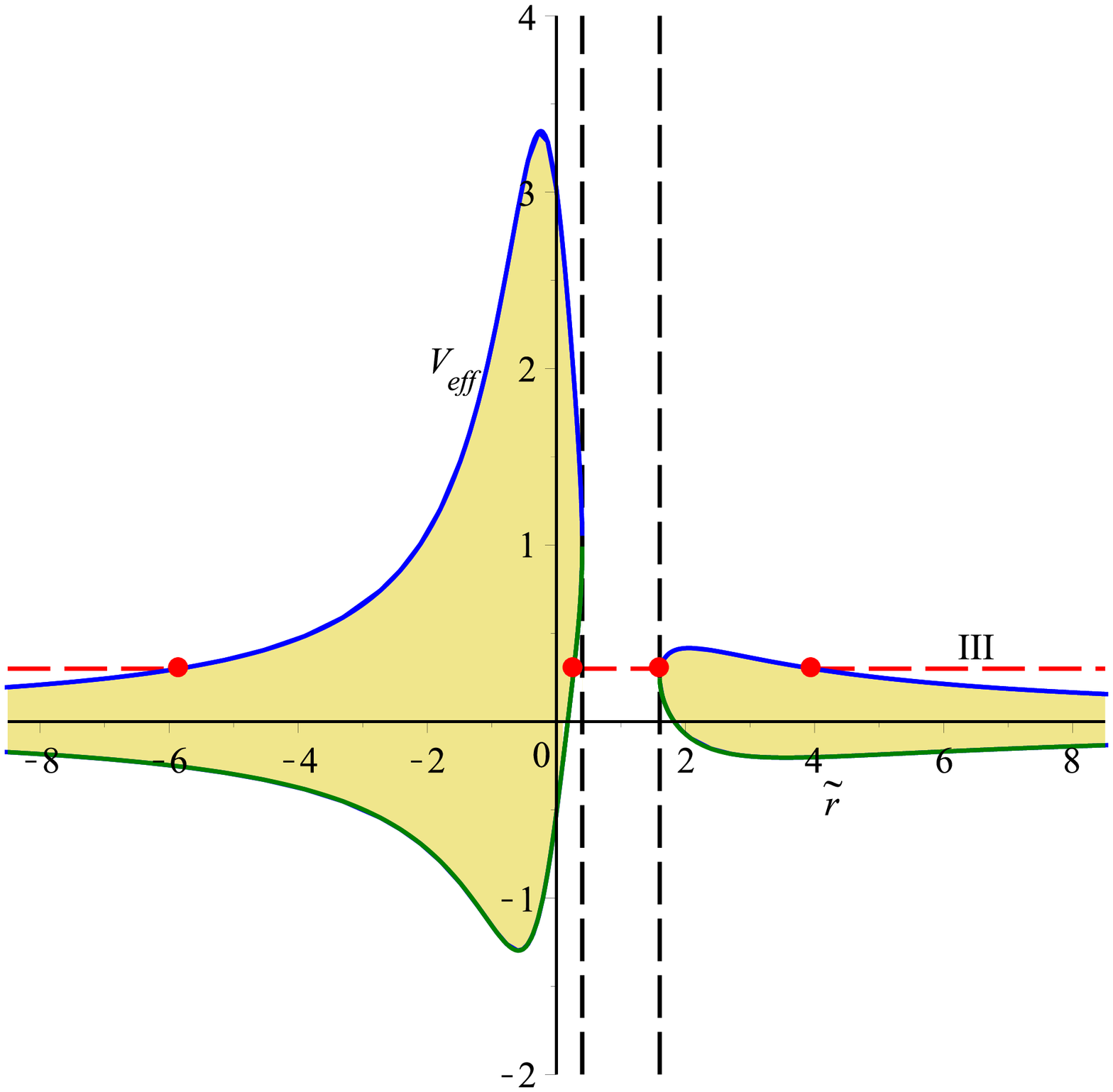}
	\label{38}
}
\hspace*{10mm}
\subfigure[ ]{
\includegraphics[width=6.5cm]{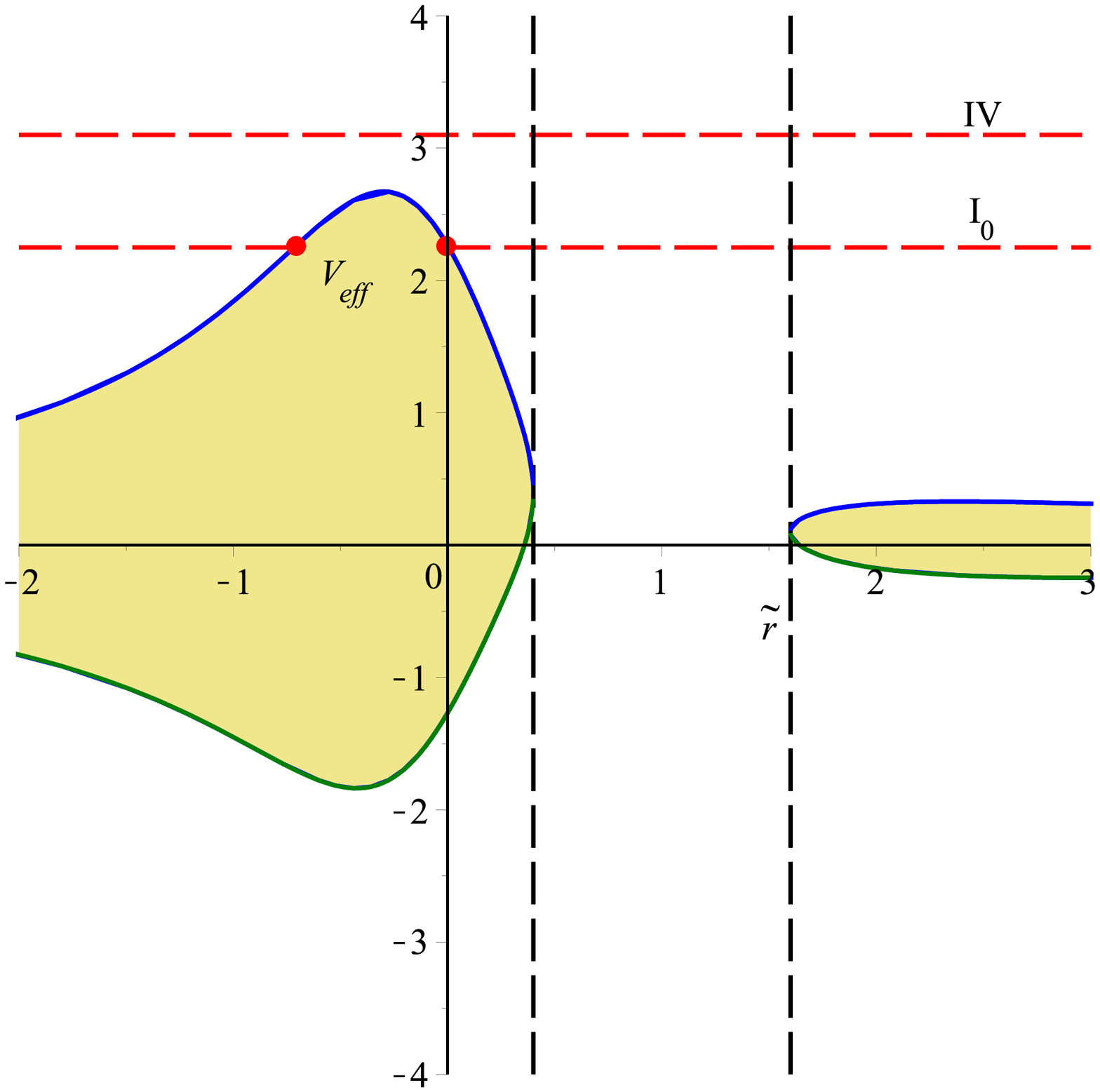}
	\label{39}
}
\caption[figs]
{Effective potentials for a rotating black string-(anti-) de sitter spacetime for $\varepsilon=1$, $\Lambda=10^{-5}$, $\tilde{a}=0.8$, $\tilde{K}=2$, $J=1$. The green curves represent to the effective potential. 
The red dashed lines denote the energy. The red dots mark the zeros of the polynomial $R$, which are the turning points of the orbits.
In the khaki area no motion is possible since $\tilde{R}<0$. The vertical black dashed lines show the position of the horizons.
  \subref{36} $\tilde{L}=1$  ، 
 \subref{37} $\tilde{L}=0.01$،
  \subref{38} $\tilde{L}=1$,
 \subref{39} $\tilde{L}=0.4$.}
\label{40}
\end{figure}

The following different types of orbits can be identified in the spacetimes described by the metric Eq.~(\ref{RBS1}), \cite{Hackmann:2010zz} and \cite{Grunau:2013oca}:
\begin{enumerate}
\item Transit orbit ($TrO$), with range $\tilde{r}$ $\in$ ($-\infty , \infty$). This orbit is a crossover orbit.
\item Escape orbit ($EO$), with range $\tilde{r}$ $\in$ [$r_{1}$,$\infty$) with $r_{1}>\tilde{r}_{+}$, or with range $\tilde{r}$ $\in$ ($-\infty$ ,$ r_{1}$] with $r_{1}<0$.
\item Two-world escape orbit ($TEO$), with range [$r_{1}, \infty$), where $0 < r_{1} < r_{-}$. The $TEOs$ cross both horizons
twice and emerge into another universe.
\item Crossover two-world escape orbit ($CTEO$), with range [$r_{1}$,$\infty$) where $r_{1} < 0$. The $CTEOs$ cross both
horizons twice and emerge into another universe. $\tilde{r} = 0$ is crossed once.
\item Bound orbit ($BO$), with range $\tilde{r} \in [r_{1}, r_{2}]$ with $0 < r_{1} < r_{2}$ and. (a) either $r_{1},r_{2}>r_{+}$ or (b) $r_{1},r_{2}<r_{-}.$
\item Many-world bound orbit ($MBO$), with range $\tilde{r} \in [r_{1}, r_{2}]$, where $0< r_{1} \leqslant r_{-}$ and $r_{2} \geqslant r_{+}$. The $MBOs$ cross both horizons several times. Each time both horizons are traversed twice the test particles emerge into another universe.
\item Terminating orbit ($TO$) with ranges $\tilde{r} \in [0,\infty)$ or $\tilde{r} \in [0, r_{1}]$ with (a) either $r_{1}\geqslant \tilde{r}_{+}$ or (b) $0<r_{1}<\tilde{r}_{-}$.
\end{enumerate}
The effective potentials related to these $\tilde{L}-E^{2}$-diagrams, whit the help of Eq.~(\ref{VR}) are shown in figs.~(\ref{40}). Also, a summary of possible orbit types  can be found in Table \ref{tab:RBsd.orbits}.
\newpage
Different regions of geodesic motion for $\tilde{\Lambda}=10^{-5}$ and $J=1$, can be identified:
\begin{enumerate}
\item ~Region I: $\tilde{R}(\tilde{r})$ has $2$ real zeros $r_{1}, r_{2}$ and $\tilde{R}(\tilde{r}) \geqslant 0$ for $\tilde{r}\in (-\infty , r_{1}]$ and $\tilde{r}\in [r_{2}, \infty)$. There is a negative
and a positive zero, so that escape orbits ($EO$) are possible for $\tilde{r}<0$ and two-world escape orbits ($TEO$) are possible for $\tilde{r}\geqslant 0$. In part $I_{0}$ there is $2$ zeros, so that escape orbits ($EO$) are possible for $\tilde{r}<0$ and the former positive zero is now at
$\tilde{r}=0$ so that the two-world escape orbit turns into a terminating orbit ($TO$).
\item ~Region II: $\tilde{R}(\tilde{r})$ has $2$ negative zeros, so there is an escape orbit ($EO$) for $\tilde{r}<0$ and a crossover two-world escape orbit ($CTEO$).
\item ~Region III: $\tilde{R}(\tilde{r})$ has $1$ negative zero $r_{1}$ and $3$ positive zero $r_{2}$, $r_{3}$, $r_{4}$. $\tilde{R}(\tilde{r})\geq 0$ for $\tilde{r}\in(-\infty , r_{1}]$, $\tilde{r}\in [r_{2}, r_{3}]$ and $\tilde{r}\in [r_{4}, \infty)$. Escape orbits ($EO$) with either $\tilde{r}<0$ or $\tilde{r}>\tilde{r}_{+}$ and many-world bound orbits ($MBO$) are possible.
\item ~Region IV: $\tilde{R}(\tilde{r})$ has $0$ real zeros and $\tilde{R}(\tilde{r})>0$ for all $\tilde{r}$. Here only transit orbits ($TrO$) are possible which cross $\tilde{r}=0$.
\end{enumerate}
\begin{table}[!ht]
\begin{center}
\begin{tabular}{|l|l|c|l|}
\hline
Region & Zeroes & Range of $\tilde{r}$ &  Types of orbits \\
\hline\hline
I & 2 &
$\textbf{---}$$\bullet$$----$$|$$--$$\bullet$$\textbf{--------}$$\Vert$$\textbf{-----------}$$\Vert$$\textbf{-----------}$

& EO, TEO
\\  \hline
II & 2 &
$\textbf{---}$$\bullet$$--$$\bullet$$\textbf{-----}$$|$$\textbf{------------}$$\Vert$$\textbf{-----------}$$\Vert$$\textbf{-----------}$
  & EO, CTEO
\\ \hline
III & 4 &
$\textbf{-----}$$\bullet$$----$$|$$--$$\bullet$$\textbf{------}$$\Vert$$\textbf{-----------}$$\Vert$$\bullet$$--$$\bullet$$\textbf{-----}$
  & EO, MBO, EO
\\ \hline
IV & 0 &
$\textbf{---------------}$$|$$\textbf{------------}$$\Vert$$\textbf{-----------}$$\Vert$$\textbf{-----------}$
  & TrO

\\ \hline
\end{tabular}
\caption{Orbit types of rotating black string-(anti) de sitter for 
$J=1$ and $ \tilde{\Lambda}=10^{-5}$. The lines represent the range of the
orbits. The turning points are shown by thick dots. The horizons are indicated by a vertical double line. The single vertical line represents $\tilde{r}=0$.}
\label{tab:RBsd.orbits}
\end{center}
\end{table}
\subsection{Analytical solution of geodesic equations}\label{ana}
In this section we introduce the analytical solution of the equations of motion Eq.~(\ref{drd})--(\ref{dtd}). Each equation will be discussed separately.
\subsubsection{\textbf{$\theta$ motion}}\label{thetam}
We start with the differential equation~(\ref{dthetad})
\begin{align}
(\dfrac{d\theta}{d\gamma})^{2}=\tilde{\Theta}(\theta)=\Delta_{\theta} \big( \tilde{K}-\varepsilon \tilde{a}^{2}\cos^{2}\theta -J^{2}\tilde{a}^{2}\cos^{2}\theta \big)-\dfrac{\chi^{2}T^2(\theta)}{\sin^{2}\theta},
\end{align}
and we substitute $\nu=\cos^{2}\theta$ to simplify the equation
\begin{align}\label{doo}
(\dfrac{d\nu}{d\gamma})^{2}=(4\nu)(1-\nu)(1+\tilde{a}^{2}\tilde{\Lambda}\nu)(\tilde{K}-\varepsilon \tilde{a}^{2}\upsilon -J^{2}\tilde{a}^{2}\nu) - (4\nu)\chi^{2}(\tilde{a}E(1-\nu)-\tilde{L})^{2}.
\end{align}

\textbf{Timelike and Null geodesics} \qquad To obtain timelike and null geodesics, the Eq.~(\ref{doo}), 
should has  simple zeros, which can be solved in terms of the Weierstrass elliptic function $\wp$. For this purpose, 
we substitute $\nu=\xi^{-1}$, and therefore 
\begin{align}\label{dkhi}
(\dfrac{d\xi}{d\gamma})^{2}=\tilde{\Theta}_{\xi},
\end{align}
where
\begin{align}\label{khiii}
\tilde{\Theta}_{\xi}=:4\xi^{3} \big[ \tilde{K}-\chi^{2}(E\tilde{a}-\tilde{L})^{2} \big] +4\xi^{2} \big[ \tilde{a}^{2}(\tilde{K}\tilde{\Lambda}-\varepsilon -J^2)-\tilde{K}+2\chi^{2}E\tilde{a}(\tilde{a}E-\tilde{L}) \big] \nonumber\\ +4\xi \tilde{a}^{2}\big[(\varepsilon +J^{2})(1-\tilde{\Lambda}\tilde{a}^{2})-E^{2}\chi^{2}-\tilde{\Lambda}\tilde{K})  \big] +4(\varepsilon +J^{2}) \tilde{\Lambda}\tilde{a}^{4} =:\sum_{i=1}^3 a_{i}\xi^{i}.
\end{align}
Now, to obtain a  Weierstrass form, we substitute $\xi=\frac{1}{a_{3}}(4y-\frac{a_{2}}{3})$, then, we have
\begin{align}\label{dy}
(\dfrac{dy}{d\gamma})^{2}=4y^{3}-g_{2}y-g_{3},
\end{align}
where
\begin{align}
g_{2}=\frac{1}{16} \big( \frac{4}{3} a_{2}^{2}-4a_{1}a_{3} \big),
\end{align}
\begin{align}\label{g3}
g_{3}=\frac{1}{16} \big( \frac{1}{3}a_{1}a_{2}a_{3}-\frac{2}{27}a_{2}^{3}-a_{0}a_{3}^{2} \big).
\end{align}
are the Weierstrass invariants. The differential equation~(\ref{dy}) represents an elliptic type, which can be solved by the Weierstrass function \cite{Hackmann:2008zz,Abramowitz:1968,Whittaker:1973}.
\begin{align}\label{yyy}
y(\gamma)=\wp \big(\gamma -\gamma_{\theta , in};g_{2},g_{3} \big).
\end{align}
So, the solution of Eq.(\ref{dthetad}) is given by
\begin{align}
\theta \big(\gamma \big) =\arccos \big( \pm \sqrt{\dfrac{a_{3}}{4\wp (\gamma -\gamma_{\theta ,in}; g_{2},g_{3})-\frac{a_{2}}{3}}} \big) ,
\end{align}
where $\gamma_{\theta ,in}=\gamma_{0}+ \int_{y_{0}}^{\infty}\dfrac{dy'}{\sqrt{4y'^{3}-g_{2}y'-g_{3}}}$ and $y_{0}=\dfrac{a_{3}}{4 \cos^{2}(\theta_{0})}+\dfrac{a_{2}}{12}$ depends only on the initial values $\gamma_{0}$ and $\theta_{0}$. Since the $\theta$ motion is symmetric with respect to the equatorial plane $\theta=\frac{\pi}{2}$, the sign of the square root can be chosen so that $\theta(\gamma)$ is either in $(0,\frac{\pi}{2})$ (positive sign) or in $(\frac{\pi}{2},\pi)$ (negative sign). 
\subsubsection{\textbf{r motion}}\label{rmotion}
The dynamics of $r$ are defined by the differential equation Eq.~(\ref{drd})
\begin{align}
(\dfrac{d\tilde{r}}{d\gamma})^{2}=\tilde{R}(\tilde{r})=\chi^{2}P^{2}(r)-\Delta_{\tilde{r}}\big(\varepsilon \tilde{r}^{2}+\tilde{K}+J^{2}\tilde{r}^{2} \big),
\end{align}
solving of this equation is more complicated because $\tilde{R}$ is a polynomial of order $6$.\\
Considering particles and light, i.e. $\varepsilon=1$, $\varepsilon=0$, and assuming that $\tilde{R}$ has only simple zeros the differential equation (\ref{drd}) is of hyperelliptic type. This equation can be solved in terms of derivatives of the Kleinian σ function, as presented in \cite{Hackmann:2008zza}. With substitution $\tilde{r}=\pm \frac{1}{u}+\tilde{r}_{\tilde{R}}$ where $\tilde{r}_{\tilde{R}}$ is a zero of $\tilde{R}$, the Eq.~(\ref{drd}) is transformed into the standard form and then we get
\begin{align}\label{udu}
\big( u\dfrac{du}{d\gamma} \big)^{2}=c_{5}\tilde{R}_{u},
\end{align}
where
\begin{align}\label{Ru}
\tilde{R}_{u}=\sum_{i=0}^{5}\frac{c_{i}}{c_{5}}u^{i}, \qquad c_{i}=\dfrac{\big( \pm 1 \big)^{i}}{(6-i)!}\dfrac{d^{(6-i)}\tilde{R}}{d u^{(6-i)}}(\tilde{r}_{\tilde{R}}).
\end{align}
We choose the sign in the substitution such that the constant $c_{5}$ is positive. We have the first kind of differential equation~(\ref{udu}) and can solve it by
\begin{align}
u(\gamma)=-\frac{\sigma_{1}}{\sigma_{2}}  \binom{f(\sqrt{c_{5}}\gamma -\gamma_{\tilde{r},in})}{\sqrt{c_{5}}\gamma - \gamma_{\tilde{r},in}},
\end{align}
here $\gamma_{\tilde{r},in}=\sqrt{c_{5}}\gamma_{0}+\int_{u_{0}}^{\infty}\dfrac{u' du'}{\sqrt{\tilde{R}_{u'}}}$ and $u_{0}=\pm (\tilde{r}_{0}-\tilde{r}_{\tilde{R}})^{-1}$ depends only on the primary values $\gamma_{0}$ and $\tilde{r}_{0}$ and  the function $f$ describes the $\theta$-divisor, i.e. $\sigma((f(x),x)^{t})=0$, eg. see, \cite{Hackmann:2008zza}. Then the solution of radial distance $\tilde{r}$ is then given by
\begin{align}
\tilde{r}(\gamma)=\mp \frac{\sigma_{2}}{\sigma_{1}} \binom{f(\sqrt{c_{5}}\gamma -\gamma_{\tilde{r},in})}{\sqrt{c_{5}}\gamma -\gamma_{\tilde{r},in}} + \tilde{r}_{\tilde{R}},
\end{align}
Here the sign depends on the sign that was chosen in the substitution $\tilde{r}=\pm \frac{1}{u}+\tilde{r}_{\tilde{R}}$, i.e. is such that $c_{5}$ in Eq.~(\ref{Ru}) is positive.
\subsubsection{\textbf{$\varphi$ motion}}\label{fii}
Solutions of the equations of motion, in this part ($\varphi$ motion) and next part ($t$ motion) are exactly similar to the solutions achieved in Ref.\cite{Hackmann:2010zz}. 
\\ 
We assume the equation for the azimuthal angle Eq.~(\ref{dphi})
\begin{align}
\dfrac{1}{\chi^{2}}(\dfrac{d\varphi}{d\gamma})=\dfrac{E(\tilde{a}^{3}+\tilde{r}^{2}\tilde{a})-\tilde{L} \tilde{a}^{2}}{\Delta_{\tilde{r}}}-\dfrac{1}{\Delta_{\tilde{\theta}}\sin^{2}\theta} \big( \tilde{a} E \sin^{2}\theta -\tilde{L} \big).
\end{align}
This equation can be separated in two parts, one only dependent on $\tilde{r}$ and the other only dependent on $\theta$. Integration yields
\begin{align}\label{phi}
\varphi - \varphi_{0}= \chi^{2} \bigg[ \int_{\gamma_{0}}^{\gamma} \dfrac{E(\tilde{a}^{3}+\tilde{r}^{2}\tilde{a})-\tilde{L} \tilde{a}^{2}}{\Delta_{\tilde{r}(\gamma)}}d\gamma - \int_{\gamma_{0}}^{\gamma}\dfrac{\tilde{a} E \sin^{2}\theta -\tilde{L}}{\Delta_{\tilde{\theta} (\gamma)}\sin^{2}\theta (\gamma)}d\gamma \bigg] \nonumber\\ =\chi^{2} \bigg[ \int_{\tilde{r}_{0}}^{\tilde{r}} \dfrac{E(\tilde{a}^{3}+\tilde{r}^{2}\tilde{a})-\tilde{L} \tilde{a}^{2}}{\Delta_{\tilde{r}}\sqrt{\tilde{R}}}d\tilde{r} - \int_{\theta_{0}}^{\theta}\dfrac{\tilde{a} E \sin^{2}\theta -\tilde{L}}{\Delta_{\tilde{\theta}}\sin^{2}\theta \sqrt{\tilde{\Theta}(\theta)}}d\theta \bigg]=\chi^{2}\bigg[ I_{r}- I_{\theta}\bigg] ,
\end{align}
Final solution of $I_{r}$ and $I_{\theta}$ yields
\begin{align}\label{akh}
I_{r}=-\dfrac{\tilde{a}u_{0}}{\sqrt{c_{5}} \mid u_{0}  \mid} \big \lbrace C_{1}(\omega -\omega_{0})+ C_{0}(f(\omega)-f(\omega_{0})) \nonumber\\ +\sum_{i=1}^{4}\dfrac{C_{2,i}}{\sqrt{\tilde{R}_{u_{i}}}} [\frac{1}{2} \log \frac{\sigma(W^{+}(\omega))}{\sigma(W^{-}(\omega))}-\frac{1}{2}\log \frac{\sigma(W^{+}(\omega_{0}))}{\sigma(W^{-}(\omega_{0}))}  \nonumber\\ -(f(\omega)-f(\omega_{0}), \omega - \omega_{0}) \big( \int_{u_{i}^{-}}^{u_{i}^{+}} d\vec{r} \big) ] \big \rbrace .
\end{align}
and
\begin{align}\label{thetafi}
I_{\theta}=\dfrac{\mid a_{3} \mid}{a_{3}}\bigg\lbrace (\tilde{a}E-\tilde{L})(\upsilon - \upsilon_{0})- \sum_{i=1}^{4}\dfrac{a_{3}}{4\chi \wp '(\upsilon_{i})} \bigg( \zeta(\upsilon_{i})(\upsilon-\upsilon_{0}) \nonumber\\+ \log \dfrac{\sigma (\upsilon -\upsilon_{i})}{\sigma(\upsilon_{0}-\upsilon_{i})} +2\pi i k_{i} \bigg) \bigg(\tilde{a}^{3}\tilde{\Lambda}(\chi E - \tilde{a}\tilde{\Lambda}\tilde{L})(\delta_{i1}+\delta_{i2}) +\tilde{L}(\delta_{i3}+\delta_{i4})\bigg) \bigg\rbrace ,
\end{align}
the details of the above solutions is given in Ref.\cite{Hackmann:2010zz}.
\subsubsection{\textbf{t motion}}
The equation for $t$ Eq.~(\ref{dtd})
\begin{align}
\dfrac{1}{\chi^{2}}(\dfrac{d\tilde{t}}{d\gamma})=\dfrac{\tilde{r}^{2}+\tilde{a}^{2}}{\Delta_{\tilde{r}}}P(r)-\dfrac{\tilde{a} }{\Delta_{\tilde{\theta}}}T(\theta) ,
\end{align}
is similar to the equation for the $\varphi$ motion. An integration yields
\begin{align}
\tilde{t}-\tilde{t}_{0}=\chi^{2} \bigg[ \int_{\gamma_{0}}^{\gamma}{\dfrac{\tilde{r}^{2}+\tilde{a}^{2}}{\Delta_{\tilde{r}}}P(r)} d\gamma - \int_{\gamma_{0}}^{\gamma} \frac{\tilde{a}}{\Delta_{\tilde{\theta}}}T(\theta) d\gamma \bigg] \nonumber\\ = \chi^{2} \bigg[ \int_{\tilde{r}_{0}}^{\tilde{r}}\dfrac{(\tilde{r}^{2}+\tilde{a}^{2})P(r)}{\Delta_{\tilde{r}}\sqrt{\tilde{R}(\tilde{r})}} d\tilde{r} - \tilde{a} \int_{\theta_{0}}^{\theta} \frac{T(\theta)}{\Delta_{\tilde{\theta}}\sqrt{\tilde{\Theta}(\theta)}}d\theta \bigg]  \nonumber\\ =\chi^{2} \bigg[ \tilde{I}_{r}-\tilde{a}\tilde{I}_{\theta} \bigg].
\end{align}
Final solution of $I_{r}$ and $I_{\theta}$ yields
\begin{align}
\tilde{I}_{r}=\dfrac{u_{0}}{\sqrt{c_{5}} \mid u_{0}  \mid} \bigg \lbrace \tilde{C}_{1}(\omega -\omega_{0})+ \tilde{C}_{0}(f(\omega)-f(\omega_{0})) \nonumber\\ +\sum_{i=1}^{4}\dfrac{\tilde{C}_{2,i}}{\sqrt{\tilde{\tilde{R}}_{u_{i}}}} \bigg[\frac{1}{2} \log \frac{\sigma(W^{+}(\omega))}{\sigma(W^{-}(\omega))}-\frac{1}{2}\log \frac{\sigma(W^{+}(\omega_{0}))}{\sigma(W^{-}(\omega_{0}))}  \nonumber\\ -(f(\omega)-f(\omega_{0}), \omega - \omega_{0}) \big( \int_{u_{i}^{-}}^{u_{i}^{+}} d\vec{r} \big) \bigg] \bigg \rbrace .
\end{align}
and
\begin{align}
\tilde{I}_{\theta}= \dfrac{\mid a_{3} \mid}{a_{3}}\bigg\lbrace(\tilde{a}E-\tilde{L})(\upsilon -\upsilon_{0})-\sum_{i=1}^{2}\dfrac{\tilde{a}a_{3}(\chi E-\tilde{a}\tilde{\Lambda}\tilde{L})}{4\wp' (\upsilon_{i})} \big[ \zeta (\upsilon_{i})(\upsilon -\upsilon_{0}) \nonumber\\ +\log\sigma (\upsilon-\upsilon_{i})-\log\sigma (\upsilon_{0}-\upsilon_{i}) \big]\bigg\rbrace,
\end{align}
the details of the above solutions is given in Ref.\cite{Hackmann:2010zz}.
\subsubsection{\textbf{w motion}}
Using the $\tilde{r}$-equation~(\ref{drd}) and the $\theta$-equation~(\ref{dthetad}), we can write the $\tilde{w}$-equation~(\ref{JJJ}) in the following way, This equation can be splitted in a part dependent only on $\tilde{r}$ and in a part only dependent on $\theta$.
\begin{align}\label{dww}
\tilde{w}-\tilde{w}_{0}=\int_{\gamma_{0}}^{\gamma}J\tilde{\rho}^{2}d\gamma =\int_{\gamma_{0}}^{\gamma}J\tilde{r}^{2}d\gamma +\int_{\gamma_{0}}^{\gamma}J\tilde{a}^{2}\cos^{2}\theta d\gamma \nonumber\\ =\int_{\tilde{r}_{0}}^{\tilde{r}}J\tilde{r}^{2}\dfrac{d\tilde{r}}{\sqrt{\tilde{R}}}+\int_{\theta_{0}}^{\theta}J\tilde{a}^{2}\cos^{2}\theta \dfrac{d\theta}{\sqrt{\tilde{\Theta}(\theta)}}.
\end{align}
We will solve now the two integrals in Eq.~(\ref{dww}) separately\\
\textbf{The $\theta$ dependent integral}\qquad Let us consider the integral
\begin{align}
I_{\theta}=\int_{\gamma_{0}}^{\gamma}J\tilde{a}^{2}\cos^{2}\theta d\gamma,
\end{align}
which can be transformed to the simpler form
\begin{align}
I_{\theta}=\pm\frac{1}{2}\int_{\nu_{0}}^{\nu}J\tilde{a}^{2}\nu \frac{d\nu}{\sqrt{\nu\tilde{\Theta}_{\nu}}},
\end{align}
by the substitution $\nu= \cos^{2}\theta$, where $d\gamma$ is defined in Eq.~(\ref{doo}).\\
So the answer of this integral can be presented:
\begin{align}
I_{\theta}=\pm\frac{1}{2}\bigg[ -\dfrac{4J\tilde{a}^{2}}{a_{3}}(\zeta_{\theta}(\upsilon)-\zeta_{\theta}(\upsilon_{0}))-\dfrac{J\tilde{a}^{2}a_{2}}{3a_{3}}(\upsilon -\upsilon_{0}) \bigg].
\end{align}
\\
\textbf{The $r$ dependent integral}\qquad
We solve now the first, $\tilde{r}$ dependent integral in Eq.~(\ref{dww})
\begin{align}
I_{r}=\int_{\gamma_{0}}^{\gamma}J\tilde{r}^{2}d\gamma,
\end{align}
with using equation~(\ref{udu}) where we substituted $d\gamma=\dfrac{u du}{\sqrt{c_{5}\tilde{R}_{u}}}$ we have:
\begin{align}\label{88}
I_{r}=\dfrac{J}{\sqrt{c_{5}}}\bigg[\int_{u_{0}}^{u}\dfrac{du}{u\sqrt{\tilde{R}_{u}}}+2\tilde{r}_{\tilde{R}}\int_{u_{0}}^{u}\dfrac{du}{\sqrt{\tilde{R}_{u}}}+\tilde{r}_{\tilde{R}}^{2}\int_{u_{0}}^{u}\dfrac{udu}{\sqrt{\tilde{R}_{u}}}\bigg].
\end{align}
The second two integrals in Eq.~(\ref{88}) are of first kind and can be expressed in terms
of $\gamma$ analogous to Sec. \ref{rmotion}, Eq.~(\ref{udu}), i.e.
\begin{align}
\int_{u_{0}}^{u}\dfrac{udu}{\sqrt{\tilde{R}_{u}}}=\sqrt{c_{5}}(\gamma -\gamma_{0}),\nonumber\\
\int_{u_{0}}^{u}\dfrac{du}{\sqrt{\tilde{R}_{u}}}=\int_{u_{0}}^{\infty}\dfrac{du}{\sqrt{\tilde{R}_{u}}}+\int_{\infty}^{u}\dfrac{du}{\sqrt{\tilde{R}_{u}}}\nonumber\\=-f(\sqrt{c_{5}}\gamma_{0}-\gamma_{\tilde{r},in})+f(\sqrt{c_{5}}\gamma -\gamma_{\tilde{r},in}),
\end{align}
where again $\gamma_{\tilde{r},in}=\sqrt{c_{5}}\gamma_{0}+\int_{u_{0}}^{\infty}\dfrac{udu}{\sqrt{\tilde{R}_{u}}}$ with $u_{0}=\pm(\tilde{r}_{0}-\tilde{R}_{\tilde{r}})^{-1}$ only depends on the initial values $\gamma_{0}$ and $u_{0}$, and $f$ describes the $\theta$-divisor,i.e. $\sigma((f(z),z)^{t})=0$.\\
The first integral in Eq.~(\ref{88}) is of third kind that containing $(u-u_{i})^{-1}$ in which we have $u_{i}=0$. This integral can be expressed in terms of the canonical integral of third kind $\int dP(x_{1},x_{2})$, In particular, we get
\begin{align}
\int_{u_{0}}^{u}\dfrac{du}{(u-u_{i})\sqrt{\tilde{R}_{u}}}=\dfrac{1}{+\sqrt{\tilde{R}_{u_{i}}}}\int_{u_{0}}^{u}dP(u_{i}^{+},u_{i}^{-}).
\end{align}
So, we have
\begin{align}
I_{r}=\dfrac{J}{\sqrt{c_{5}}}\bigg\lbrace \tilde{r}_{\tilde{R}}^{2} \sqrt{c_{5}}(\gamma -\gamma_{0})+2\tilde{r}_{\tilde{R}}\bigg[ f(\sqrt{c_{5}}\gamma_{0}-\gamma_{\tilde{r},in})+f(\sqrt{c_{5}}\gamma -\gamma_{\tilde{r},in})\bigg] \nonumber\\+\dfrac{1}{\sqrt{\tilde{R}_{u_{i}}}}\bigg[\frac{1}{2}\log\dfrac{\sigma(W^{+}(\omega))}{\sigma(W^{-}(\omega))}-\frac{1}{2}\log\dfrac{\sigma(W^{+}(\omega_{0}))}{\sigma(W^{-}(\omega_{0}))} \nonumber\\ -(f(\omega)-f(\omega_{0}), \omega - \omega_{0}) \big( \int_{u_{i}^{-}}^{u_{i}^{+}} d\vec{r} \big)\bigg]\bigg\rbrace .
\end{align}
The final solution of the $\tilde{w}$-equation is
\begin{align}
\tilde{w}=\pm\frac{1}{2}\bigg[ -\dfrac{4J\tilde{a}^{2}}{a_{3}}(\zeta_{\theta}(\upsilon)-\zeta_{\theta}(\upsilon_{0}))-\dfrac{J\tilde{a}^{2}a_{2}}{3a_{3}}(\upsilon -\upsilon_{0}) \bigg] + \dfrac{J}{\sqrt{c_{5}}}\bigg\lbrace \tilde{r}_{\tilde{R}}^{2} \sqrt{c_{5}}(\gamma -\gamma_{0}) \nonumber\\ +2\tilde{r}_{\tilde{R}}\bigg[ f(\sqrt{c_{5}}\gamma_{0}-\gamma_{\tilde{r},in})+f(\sqrt{c_{5}}\gamma -\gamma_{\tilde{r},in})\bigg] +\dfrac{1}{\sqrt{\tilde{R}_{u_{i}}}}\bigg[\frac{1}{2}\log\dfrac{\sigma(W^{+}(\omega))}{\sigma(W^{-}(\omega))} \nonumber\\ -\frac{1}{2}\log\dfrac{\sigma(W^{+}(\omega_{0}))}{\sigma(W^{-}(\omega_{0}))}  - (f(\omega)-f(\omega_{0}), \omega - \omega_{0}) \big( \int_{u_{i}^{-}}^{u_{i}^{+}} d\vec{r} \big)\bigg]\bigg\rbrace +\tilde{w}_{0} .
\end{align}
\subsection{orbits}
\begin{figure}[!ht]
\centering
\subfigure[ ]{
\includegraphics[width=8cm]{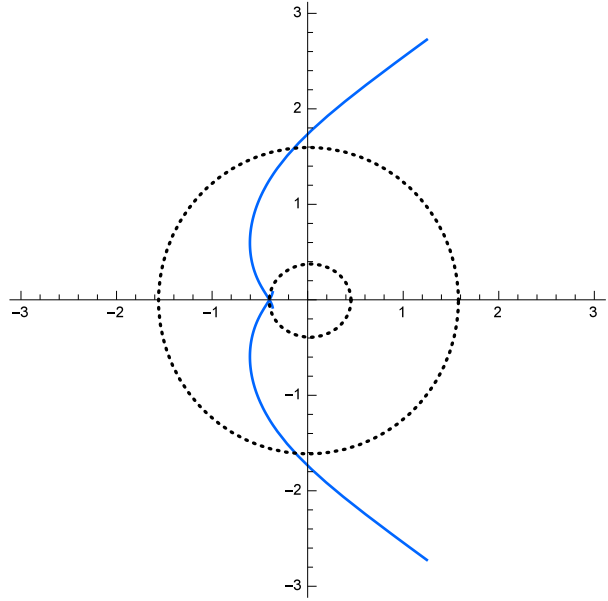}\label{41}}

\caption[figs]
{Two-world escape orbit (TEO) in Region I with, $\varepsilon=1$, $\tilde{a}=0.8$, $\tilde{K}=2$, $\tilde{\Lambda}=10^{-5}$, $J=1$, $E=\sqrt{0.5}$, $\tilde{L}=1$.}
\end{figure}

\begin{figure}[!ht]
\centering
\subfigure[ ]{
\includegraphics[width=8cm]{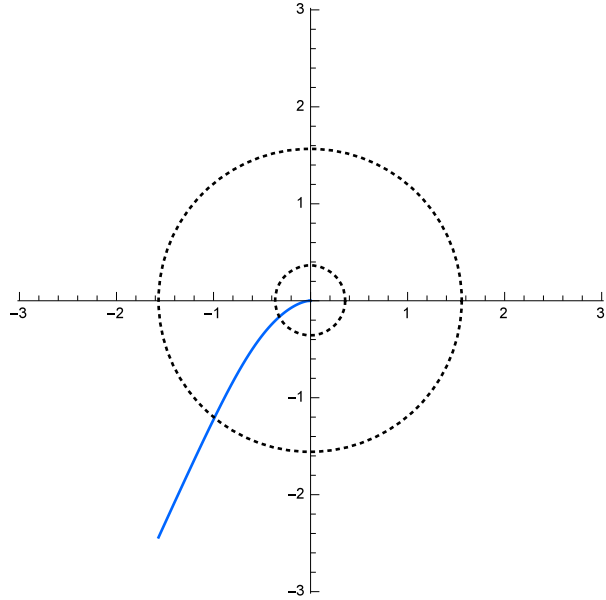}\label{42}}
\caption[figs]
{Terminating orbit (TO) in Region $I_{0}$ with, $\varepsilon=1$, $\tilde{a}=0.8$, $\tilde{K}=2$, $\tilde{\Lambda}=10^{-5}$, $J=1$, $E=\sqrt{5.14}$, $\tilde{L}=0.4$.}
\end{figure}

\begin{figure}[!ht]
\centering
\subfigure[ ]{
\includegraphics[width=8cm]{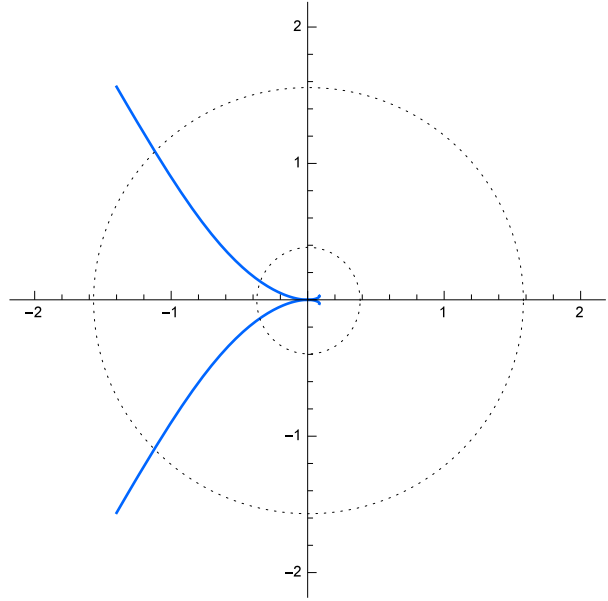}\label{43}}
\caption[figs]
{Crossover two-world escape orbit (CTEO) in Region II with, $\varepsilon=1$, $\tilde{a}=0.8$, $\tilde{K}=2$, $\tilde{\Lambda}=10^{-5}$, $J=1$, $E=\sqrt{4}$, $\tilde{L}=0.01$.} 
\end{figure}

\begin{figure}[!ht]
\centering
\subfigure[ ]{
\includegraphics[width=8cm]{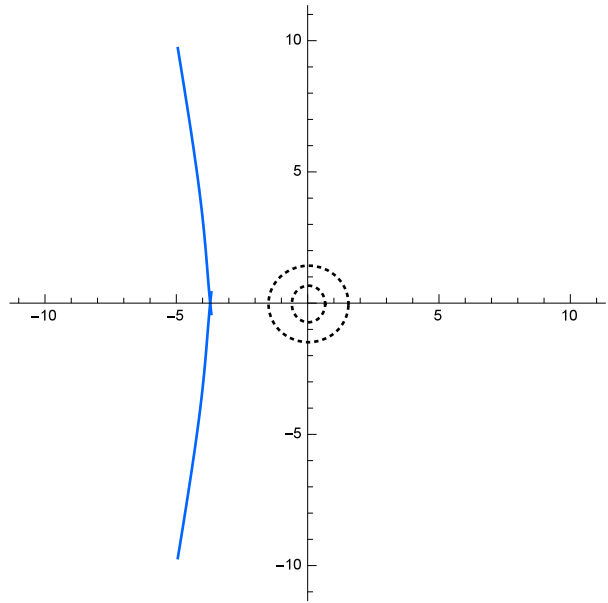}\label{44}}
\caption[figs]
{Escape orbit (EO) in Region III with, $\varepsilon=1$, $\tilde{a}=0.8$, $\tilde{K}=2$, $\tilde{\Lambda}=10^{-5}$, $J=1$, $E=\sqrt{0.1}$, $\tilde{L}=1$.} 
\end{figure}

\begin{figure}[!ht]
\centering
\subfigure[ ]{
\includegraphics[width=8cm]{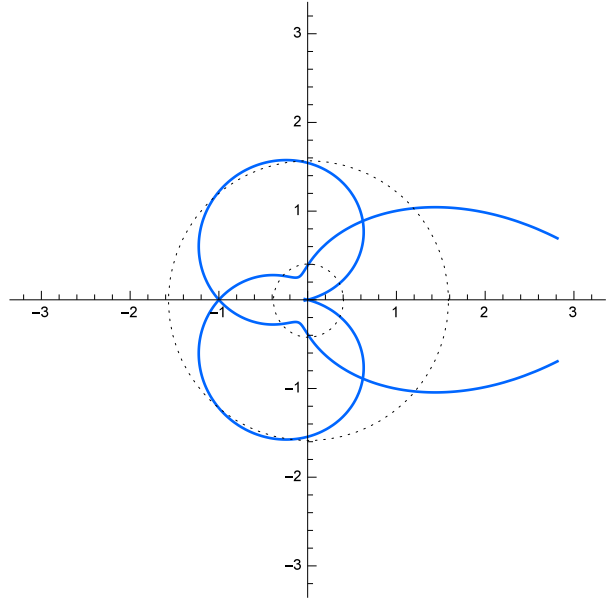}\label{45}}
\caption[figs]
{Many-world bound orbit (MBO) in Region III, with $\varepsilon=1$, $\tilde{a}=0.8$, $\tilde{K}=2$, $\tilde{\Lambda}=10^{-5}$, $J=1$, $E=\sqrt{0.1}$, $\tilde{L}=1$.} 
\end{figure}

\begin{figure}[!ht]
\centering
\subfigure[ ]{
\includegraphics[width=8cm]{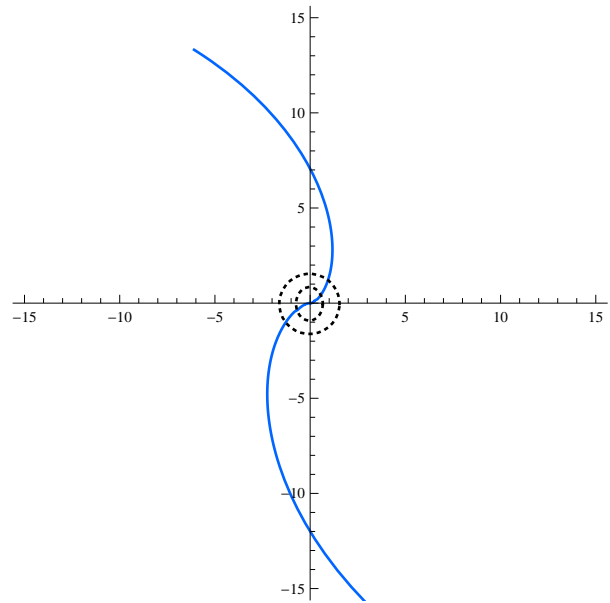}\label{46}}
\caption[figs]
{Transit orbit (TrO) in Region IV with $\varepsilon=1$, $\tilde{a}=0.8$, $\tilde{K}=2$, $\tilde{\Lambda}=10^{-5}$, $J=1$, $E=\sqrt{9}$, $\tilde{L}=0.4$.} 
\end{figure}

In this part, we plot examples of possible orbit types of geodesic motions for each region with the help of obtained analytical solutions, parametric 
$ \tilde{L}-E^{2} $- diagrams (Fig. \ref{27}--~\ref{35}) and effective potentials (Fig.~\ref{40}). These orbits are demonstrated in Figs. \ref{41}--~\ref{46}. It can be seen from Fig.~\ref{41}, for given parameter in regain I, we have a TEO, motion, where the test particle cross both horizons twice and emerge into another universe. Example of TO, motion in region $I_{0}$ is shown in Fig.~\ref{42}, where, the test particle comes from infinity and fall into center of black hole.
Also, it can be observed from Fig.~\ref{43}, in region II, we have a CTEO motion, in which, the test particle crosses both horizons twice and also crosses $\tilde{r}=0$ once, then, emerge into another universe.
In, region III, we have a EO, motion, in which, the test particle comes from infinity, close to black hole and goes to infinity again.
Moreover, example of MBO, motion for region III is shown Fig.~\ref{45}, where the test particle, crosses both horizons many times and emerge into another universe again.
Final example is TrO motion in region IV, in which, the test particle, comes from infinity, crosses horizons and goes to infinity again.

\newpage

\newpage

\clearpage

\section{Conclusion}
In this article we derived the analytical solutions of the geodesic equations for both test
particles and light rays in the both static and the rotating black string-(anti-) de sitter spacetime. 
By adding a compact dimension to the Schwarzschild-(anti-) de sitter and Kerr-(anti-) de sitter metric, the static and the rotating black string-(anti-) de sitter metric are acquired. 
The analytical expressions for the orbits ($r, \theta, \varphi, t, w$) are given by elliptic Weierstrass and hyperellliptic Kleinian functions. We classified possible types of geodesic motion by 
an analysis of the zeros of the polynomials underlying the $\theta$ and $\tilde{r}$ motion. Using effective potential  techniques and parametric diagrams, possible types of orbits were derived. In the static case EO, BO,TO and TO are possible, while in the rotating case BO, MBO, EO, TEO, CTEO, TrO and TO are possible.

\bibliographystyle{amsplain}

\end{document}